\documentclass[5p,times,twocolumn]{elsarticle}
\usepackage{lineno,hyperref}
\usepackage{subcaption}
\usepackage{tikz}
\usepackage{pgfplotstable}
\usepackage{pgfplots}
\usepackage{multirow}
\usepackage{float}
\usepackage{textcomp}
\usepackage{amsmath,amsfonts,amssymb,bbm}	
\usepackage{multicol}
\usepackage{soul}
\usepackage{graphicx}

\usepackage{algorithm}
\usepackage[noend]{algpseudocode}
\usepackage{multicol}
\usepackage{xcolor}
\usepackage{tcolorbox}

\usepackage[section]{placeins}

\usepackage{blindtext}
\usepackage{booktabs}

\makeatletter
\def\ps@pprintTitle{%
    \let\@oddhead\@empty
    \let\@evenhead\@empty
    \def\@oddfoot{\reset@font\hfil\thepage\hfil}
    \let\@evenfoot\@oddfoot
}
\makeatother

\modulolinenumbers[5]

\biboptions{authoryear}

\begin{document}

\begin{frontmatter}

\title{Weakly-supervised segmentation using inherently-explainable classification models and their application to brain tumour classification}

\author[1,2,3]{Soumick Chatterjee\corref{equalcontribution}}
\author[4,5,6]{Hadya Yassin\corref{equalcontribution}}
\author[7]{Florian Dubost}
\author[1,7]{Andreas N{\"u}rnberger}
\author[2,7,9]{Oliver~Speck}

\cortext[equalcontribution]{S. Chatterjee and H. Yassin have Equal Contribution}

\address[1]{Data and Knowledge Engineering Group, Faculty of Computer Science, Otto von Guericke University Magdeburg, Germany}
\address[2]{Biomedical Magnetic Resonance, Faculty of Nature Sciences, Otto von Guericke University Magdeburg, Germany}
\address[3]{Genomics Research Centre, Human Technopole, Milan, Italy}
\address[4]{Institute for Medical Engineering, Faculty of Electrical Engineering and Information Technology, Otto von Guericke University Magdeburg, Germany}
\address[5]{Digital Engineering Faculty, University of Potsdam, Germany}
\address[6]{Digital Health and Machine Learning, Hasso-Plattner-Institute, Potsdam, Germany}
\address[7]{Department of Biomedical Data Science, Stanford University, Stanford, CA, United States}
\address[8]{Centre for Behavioural Brain Sciences, Magdeburg, Germany}
\address[9]{German Centre for Neurodegenerative Disease, Magdeburg, Germany}

\begin{abstract}
Deep learning has demonstrated significant potential in medical imaging; however, the opacity of "black-box" models hinders clinical trust, while segmentation tasks typically necessitate laborious, hard-to-obtain pixel-wise annotations. To address these challenges simultaneously, this paper introduces a framework for three inherently explainable classifiers (GP-UNet, GP-ShuffleUNet, and GP-ReconResNet). By integrating a global pooling mechanism, these networks generate localisation heatmaps that directly influence classification decisions, offering inherent interpretability without relying on potentially unreliable post-hoc methods. These heatmaps are subsequently thresholded to achieve weakly-supervised segmentation, requiring only image-level classification labels for training. Validated on two datasets for multi-class brain tumour classification, the proposed models achieved a peak F1-score of 0.93. For the weakly-supervised segmentation task, a median Dice score of 0.728 (95\% CI: 0.715-0.739) was recorded. Notably, on a subset of tumour-only images, the best model achieved an accuracy of $98.7\%$, outperforming state-of-the-art glioma grading binary classifiers. Furthermore, comparative Precision-Recall analysis validated the framework's robustness against severe class imbalance, establishing a direct correlation between diagnostic confidence and segmentation fidelity. These results demonstrate that the proposed framework successfully combines high diagnostic accuracy with essential transparency, offering a promising direction for trustworthy clinical decision support.
\end{abstract}


\end{frontmatter}


\section{Introduction}
Convolutional neural networks (CNN) are widely used in image processing, including in the medical field. Although CNNs conducted in a supervised manner have shown significant outcomes in disease classification tasks, traditional means are still the go-to method for diagnosis, such as histopathological analysis of biopsy specimens to classify and diagnose brain tumours~\citep{Noor.2019,Irmak.2021}. A biopsy procedure is invasive, time-consuming, and susceptible to manual errors. Early detection of tumours is crucial for the survival of the patient, and therefore an automated deep-learning method might offer a better solution and overcome the disadvantages above. While early detection and appropriate diagnosis are critical for a patient's survival, segmenting the tumour for further inspection and monitoring of the disease's progression, as well as aiding in the treatment process, is also necessary.

Some critical issues must be addressed before deep learning methods can be widely accepted and adopted in the medical field. Because of the complexity of the CNNs reasoning process, they tend to be obscure and are frequently referred to as black boxes. While some researchers came up with post-hoc methods to explain and interpret such models \citep{chatterjee2021torchesegeta}, \citep{KENNY2021103459}, \citep{laugel2020local}, \citep{DBLP:journals/corr/abs-2108-04840}, others try to create inherently explainable-interpretable models. Additionally, most deep learning based segmentation models are based on supervised learning and require manually annotated data, which is time-consuming and imposes a big challenge. Nevertheless, labelling the data with the type of tumour or whether it is present or not is much easier for radiologists and, hence, less time-consuming. \citet{Dubost.5222017} has shown the potential of training a segmentation model in a weakly-supervised manner using regression labels for lesion counts. This research aims to develop fully automated, inherently explainable CNN models for brain tumour multi-class classification, which are also capable of tumour segmentation without the need for segmentation labels. 
\subsection{Related Work}
\label{sec:related_work}
Deep learning architectures have evolved significantly, demonstrating versatility across diverse image processing domains beyond medical imaging. For instance, encoder-decoder networks and adaptive feature fusion have shown great success in complex image restoration and inpainting tasks \cite{chen2025dual,zhang2025atm}. Similarly, cross-modality mutual guidance and attention mechanisms have been effectively leveraged to enhance feature representation in multi-modal object tracking \cite{zhang2025mgnet,zhang2025rgbt}. Deep learning has revolutionised diagnostics across various organs, including breast cancer \cite{singh2025cicada,singh2025comprehensive}, diabetic retinopathy \cite{singh2025comprehensiveratino,banerjee2025advances}, and lung cancer \cite{banerjee2025towards}, often employing attention mechanisms and explainable AI strategies \cite{banerjee2025comparing,banerjee2025electromagnetic}.

In the medical domain, these architectural advancements have been adapted to address brain tumour analysis. Several models have been proposed for the specific tasks of classification and segmentation, while some address both problems in a unified fashion. 

\subsubsection{Classification Models}
\label{sec:related_work_classify}
Without any prior region-based segmentation, \citet{Abiwinanda.2019} applied the most straightforward possible architecture of CNN on the brain tumour dataset with a validation accuracy of 84.19\% at best. \citet{Sultan.2019} proposed a CNN architecture for classifying brain cancers into the three classes of tumours in the brain tumour dataset, as well as distinguishing between the three glioma categories (Grade II, Grade III, and Grade IV). \citet{Badza.2020} created a 22-layered CNN architecture for brain tumour type classification. \citet{Ayadi.2021} proposed a CNN-based computer-assisted diagnosis (CAD) system using an 18-weighted layered CNN model for brain classification on two distinct datasets.  

As for models based on transfer learning, \citet{Khawaldeh.2018} developed a modified version of the AlexNet CNN model to classify brain MRI images into healthy, low-grade glioma and high-grade glioma. \citet{Deepak.2019} employed a pre-trained GoogleNet CNN model to distinguish between the three types of tumours. \citet{Rehman.2020} proposed using three pre-trained CNN models known as AlexNet, GoogleNet, and VGG16 to classify tumours in the brain tumour dataset. Building upon these transfer learning strategies, recent studies have integrated attention mechanisms to enhance clinical reliability; for instance, \citet{banerjee2025pyramidal} proposed a pyramidal attention-based network to refine feature extraction and improve classification performance. Furthermore, as trust in automated diagnosis becomes paramount, the focus has shifted towards interpretable architectures. Recent works have proposed explainable CNNs for automated detection \cite{hasan2023explainable} and case-based interpretable models like MProtoNet \cite{wei2024mprotonet}, which utilise multi-parametric MRI to provide transparent reasoning alongside classification results. Complementing these attention-based strategies, recent research has also focused on enhanced feature integration; for example, the T-FSPANNet model employs a tri-attribute and pyramidal attention-based feature fusion approach to ensure accurate and interpretable brain tumour diagnoses \cite{pacal2026towards}

\subsubsection{Segmentation Models}
\label{sec:related_work_seg}
\citet{Pereira.2016} developed a 3x3 kernels-based CNN and yielded 88\%, 83\%, 77\% dice coefficients in the BraTS 2013 challenge for complete, core, and enhancing tumour, respectively. They won second place in the BraTs 2015 challenge with a Dice coefficient of 78\%, 65\%, and 75\%. Razzak et al. (2018) presented a new model two-pathway-group CNN architecture for brain tumour segmentation that concurrently uses local and global contextual data. They include the cascade design into a two-pathway-group CNN, in which the output of a basic CNN is considered a second source and concatenated at the final layer. Cascade CNN was Superior to the CNN alone and yielded 88.9\%, 81.1\%, 73.7\% dice coefficients on the BraTS 2013 dataset for complete, core, and enhancing tumours, respectively. They also achieved 89.2\%, 79.18\%, 75.1\% dice coefficients on the BraTS 2015 dataset\citep{Razzak.2018}. Li et al. (2019) proposed a novel CNN based on UNet, which improves information flow in the network by using a unique structure called an up-skip connection between the encoding and decoding paths. Furthermore, each block includes an inception module to aid the network in learning richer representations in addition to effective cascade training techniques for successively segmenting brain tumour subregions. The model scores a dice value of 89\% for the complete tumour of the BraTS 2017 dataset\citep{Li.2019}. Similar hybrid strategies have been successfully validated in other domains, such as the Inception-U-Net gravitational optimisation model (UIGO) used for liver tumour detection \cite{banerjee2025UIGO}, confirming the utility of combining Inception modules with U-Net architectures. Beyond neuro-oncology, the efficacy of hybrid and attention-driven architectures has been validated across various anatomical segmentation tasks. Novel approaches combining deep feature attention \cite{banerjee2025novel} and multi-attention mechanisms \cite{narayan2025comparative} have demonstrated that integrating statistical validation and complex attention blocks significantly improves segmentation accuracy in challenging medical imaging contexts.

\subsubsection{Combined Models}
\label{sec:related_work_combined}
 Dubost et al. (2017) developed a novel convolutional neural network, which serves as the baseline for this work. The network is an UNet based 3D regression network that detects lesions from weak labels. They achieved a sensitivity of 62\% with an average of 1.5 false positives per image by combining a lesion count regression task with lesion detection in one model, where only a single label per image (lesion count) is required for training\citep{Dubost.5222017}. \citet{DiazPernas.2021} developed a tumour classification and segmentation model based on pixel-wise classification into four classes, including healthy, meningioma, glioma, and pituitary tumour. This method was trained and tested on the brain tumour dataset achieved 5-fold cross-validation average values of 94\% for classification sensitivity and 0.82 for segmentation dice coefficient. Additionally, they scored a 97.3\% classification accuracy. More recently, the field has progressed towards utilising Transformer-based architectures for this task. Novel approaches such as WS-MTST \cite{chen2023ws} and channel-gated Transformers with Affinity Class Activation Mapping (CAM) \cite{han2025channel} have been introduced to tackle weakly supervised multi-class segmentation, leveraging self-attention mechanisms to refine localisation from weak labels.

\subsection{Motivation and Contributions}
\label{sec:motivate}
Despite the remarkable performance of different deep learning models in medical imaging, the "black-box" nature of these models remains a significant barrier to clinical trust and adoption. While post-hoc interpretability methods, such as occlusion or gradient-based techniques, have been developed to approximate model reasoning, their fidelity is often debated. These methods attempt to explain a decision after it has been made, rather than revealing the decision-making process itself, potentially leading to discrepancies between the explanation and the model's actual focus. Consequently, there is a critical clinical need for inherently explainable models where the mechanism of localisation is integral to the classification task itself, ensuring that the features driving the diagnosis are transparent to the clinician. Furthermore, the development of robust segmentation models traditionally relies on fully supervised learning, which necessitates pixel-wise manual annotations. Obtaining these ground-truth segmentation masks is an arduous, time-consuming process requiring high-level radiological expertise, which creates a significant bottleneck in data preparation. In contrast, image-level classification labels (e.g. "tumour present" or "tumour type") are far easier and cheaper to acquire. Finally, whilst recent methodological advancements have seen the proliferation of heavy architectures such as Vision Transformers (ViTs), the translation of such models into routine clinical practice remains constrained by their significant computational overhead. The quadratic complexity inherent to self-attention mechanisms necessitates high-performance computing infrastructure that is frequently unavailable in standard medical facilities. Consequently, this reliance on substantial memory and processing power renders these architectures largely unsuitable for deployment in resource-constrained environments, where algorithms must be efficient enough to operate on consumer-grade hardware without compromising diagnostic accuracy.

To address these challenges, this work introduces a versatile framework for inherently explainable classification and weakly-supervised segmentation. The primary contributions of this paper are as follows:
\begin{itemize}
    \item \textbf{Inherently Explainable Framework:} This paper proposes a novel classification framework that is explainable by design. By integrating a Global Pooling (GP) mechanism, the network’s focus areas (localisation heatmaps) are directly used for decision-making, eliminating the need for unreliable post-hoc interpretation tools.
    \item \textbf{Versatile Backbone Integration:}The elegance and flexibility of the framework are demonstrated by showing that various distinct architectures can be successfully converted into explainable classifiers. Experiments were conducted using UNet (a standard segmentation network), as well as ReconResNet and ShuffleUNet (reconstruction and super-resolution networks), proving the framework's applicability across different domain-specific backbones.
    \item \textbf{Weakly-Supervised Segmentation:} Explainability heatmaps are post-processed here into binary segmentation masks using thresholding techniques. This allows the models to perform tumour segmentation despite being trained solely with image-level classification labels, significantly reducing the annotation burden.
    \item \textbf{Resource Efficiency:} Unlike heavy 3D models or complex Transformer architectures, the proposed 2D GP-models are lightweight and computationally efficient. 
\end{itemize}

\section{Methodology}

\subsection{Network Models}
\begin{figure*}   
\centering
\includegraphics[width=\textwidth]{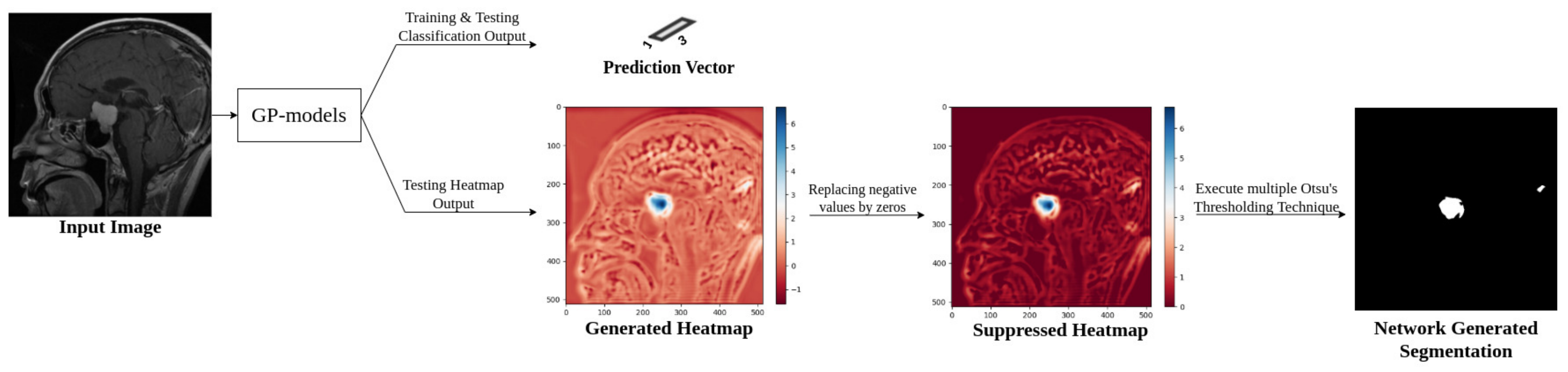}
\caption[Workflow of the GP-models]{Workflow of the GP-models}
\label{fig:Approach}
\end{figure*}

The proposed CNNs are based on modified segmentation or reconstruction models - modified to work as a multi-class classifier with only one global class label per image. The original models generate $n$ equally sized output images with the same size as the input, where $n$ is the number of segmentation classes that were required. In this proposed approach, prior to applying the final fully-connected convolution layer, the modified CNN network is coupled with a global pooling layer in the training stage to aggregate the 3D output into a single-pixel (neuron) for each class. As a result, classification image-level labels work for training the model in a supervised manner, as an alternative to the traditional cumbersome segmentation labels during the network's training - making the segmentation training of the network weakly-supervised. During the testing stage, the global pooling layer can be removed to obtain the heatmaps with the same size as the input, or it can be kept as in the training stage to obtain the class. The obtained heatmaps show the region-of-interest of the network - portraying the network's focus area - making the network inherently explainable.
The resulting heatmaps are then suppressed by setting the negative values to zero. Without using any post-hoc interpretability methods, the model's focus areas can be understood simply by glancing at the heatmaps. Afterwards, the segmentation mask is created using Otsu thresholding. The general overview of the models is shown in Fig.~\ref{fig:Approach}and Algo. \ref{alg:gp_models} presents a  pseudocode describing the training and inference procedures for the GP-Models framework.

\begin{algorithm*}
\caption{GP-Models: Supervised Training and Weakly-supervised Segmentation}
\label{alg:gp_models}
\begin{algorithmic}[1]
\State \textbf{Input:} Training Dataset $\mathcal{D} = \{(X_i, y_i)\}_{i=1}^{N}$ where $X$ is an MRI slice and $y$ is the class label.
\State \textbf{Input:} Backbone Network $\Phi$ (e.g. ReconResNet or UNet or ShuffleUNet) producing feature maps $F$.
\State \textbf{Input:} Classifier Layer $\Psi$ ($1\times1$ Convolution).
\State \textbf{Hyperparameters:} Learning rate $\alpha$, Epochs $E$.

\Statex
\Function{TrainModel}{$\mathcal{D}, \Phi, \Psi$}
    \State Initialise weights $\theta_{\Phi}$ and $\theta_{\Psi}$.
    \For{epoch $e \in \{1, \dots, E\}$}
        \For{batch $(X_b, y_b) \in \mathcal{D}$}
            \State \Comment{\textbf{Phase 1: Feature Extraction}}
            \State $F_b \gets \Phi(X_b)$ \Comment{Output shape: $[Batch, Channels, H, W]$}
            
            \State \Comment{\textbf{Phase 2: Global Pooling (The GP-Module)}}
            \State $v_b \gets \text{GlobalMaxPooling}(F_b)$ \Comment{Output shape: $[Batch, Channels, 1, 1]$}
            
            \State \Comment{\textbf{Phase 3: Classification}}
            \State $\hat{y}_b \gets \Psi(v_b)$ \Comment{Apply $1\times1$ Conv to vector}
            
            \State \Comment{\textbf{Optimisation}}
            \State $\mathcal{L} \gets \text{CrossEntropyLoss}(\hat{y}_b, y_b)$
            \State Update $\theta_{\Phi}, \theta_{\Psi}$ using gradient descent on $\nabla \mathcal{L}$
        \EndFor
    \EndFor
    \State \Return Trained $\Phi, \Psi$
\EndFunction

\Statex
\Function{GenerateSegmentation}{$X_{test}, \Phi, \Psi$}
    \State \Comment{Inference for Explainability/Segmentation}
    
    \State $F_{test} \gets \Phi(X_{test})$ \Comment{Extract spatial features}
    
    \State \Comment{\textbf{Bypass Global Pooling}}
    \State $H_{raw} \gets \Psi(F_{test})$ \Comment{Apply $1\times1$ Conv spatially to get Heatmap}
    \State \Comment{$H_{raw}$ shape is $[1, NumClasses, H, W]$}
    
    \State \Comment{\textbf{Post-Processing}}
    \State $H_{supp} \gets \max(0, H_{raw})$ \Comment{Suppress negative activations (ReLU)}
    
    \State $S_{mask} \gets \text{zeros\_like}(H_{supp})$
    \For{class $c \in Classes$}
        \State $\tau \gets \text{OtsuThreshold}(H_{supp}[c])$ \Comment{Calculate dynamic threshold}
        \State $S_{mask}[c] \gets H_{supp}[c] > \tau$ \Comment{Binarise the heatmap}
    \EndFor
    
    \State \Return $H_{supp}$ (Heatmap), $S_{mask}$ (Segmentation)
\EndFunction

\end{algorithmic}
\end{algorithm*}

\begin{figure*}   
\centering
\includegraphics[width=\textwidth]{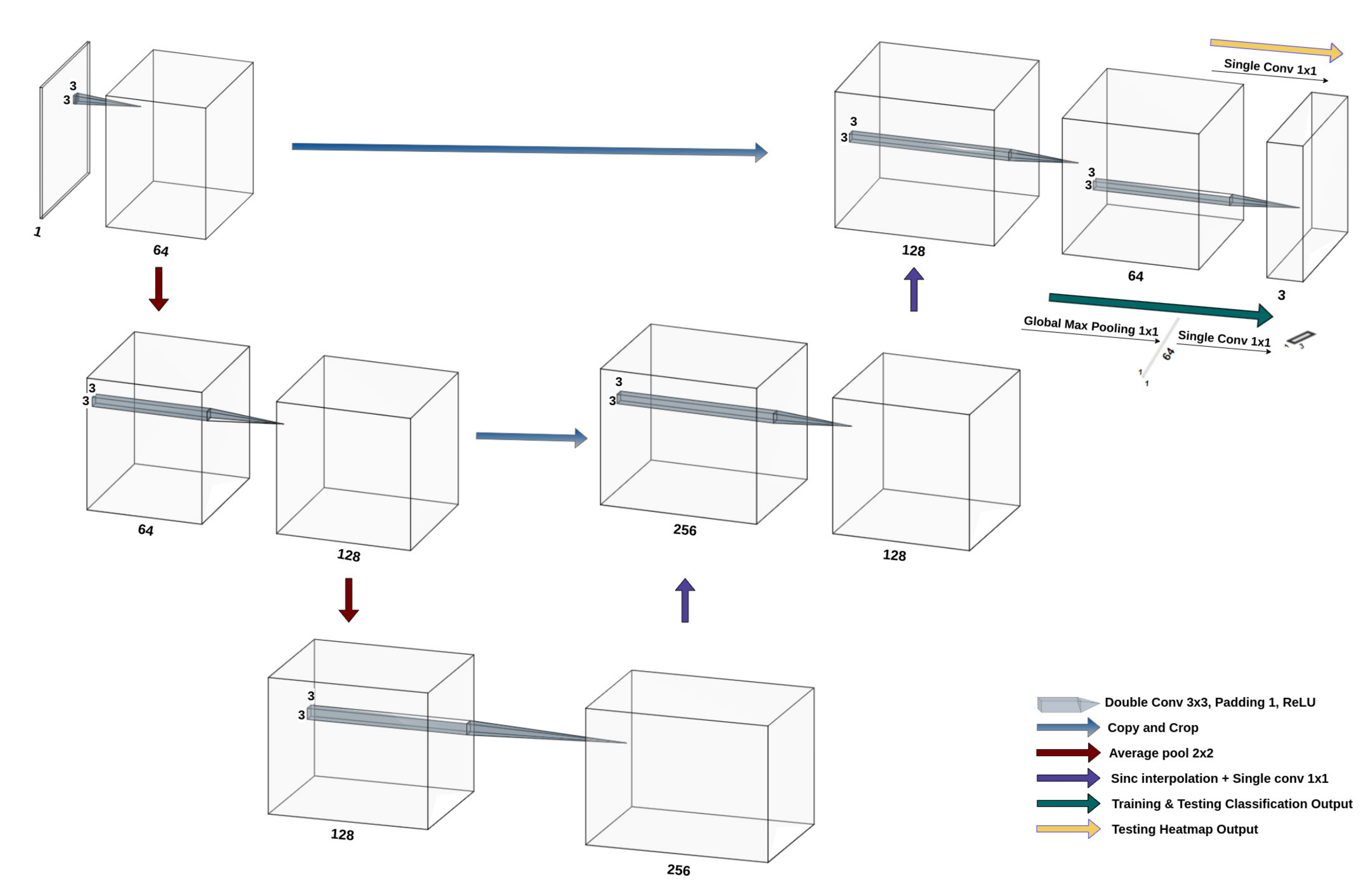}
\caption{The original Network Architecture of the baseline GP-UNet model\citep{Dubost.5222017} was modified by changing the up-pooling mechanism from transposed-convolution to interpolation+convolution (Sinc Up-sample method) and changing the output convolution filter number from $2^{5+i}$ to $2^{6+i}$, $i=0, 1, 2$ corresponding to the original depth of the model (depth=3). A dropout layer with a probability value of 0.5 was added to the model during training.}
\label{fig:GP-UNet}
\end{figure*}

\begin{figure*}   

\centering
\includegraphics[width=\textwidth]{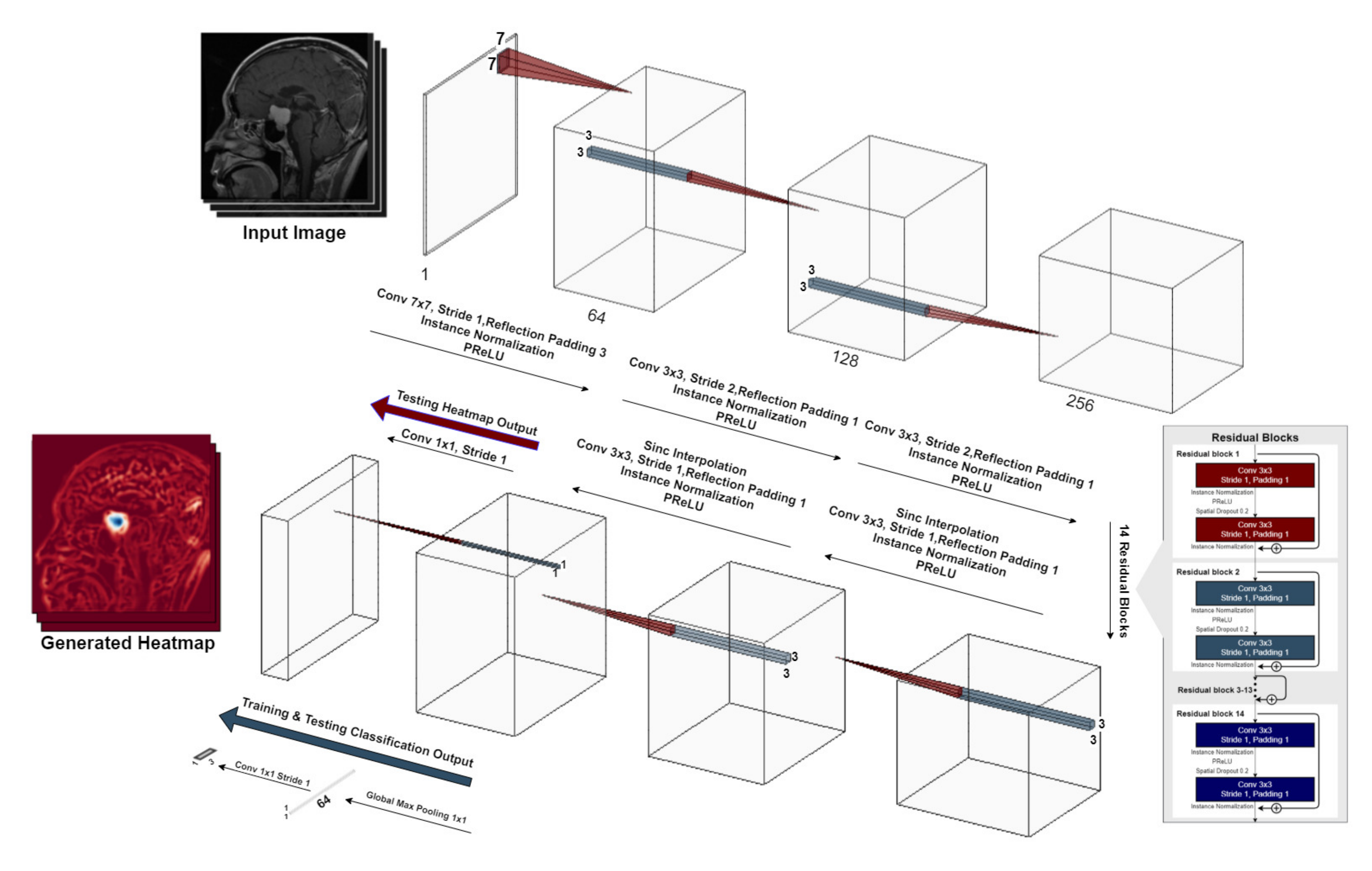}
\caption[The Network Architecture of the proposed GP-ReconResNet model.]{The Network Architecture of the proposed GP-ReconResNet model. When training and testing on the BraTS 2020 dataset, a dropout layer with a probability value of 0.5 instead of 0.2 was used in the model.}
\label{fig:Recon}
\end{figure*}

\begin{figure*}   
\centering
\includegraphics[width=\textwidth]{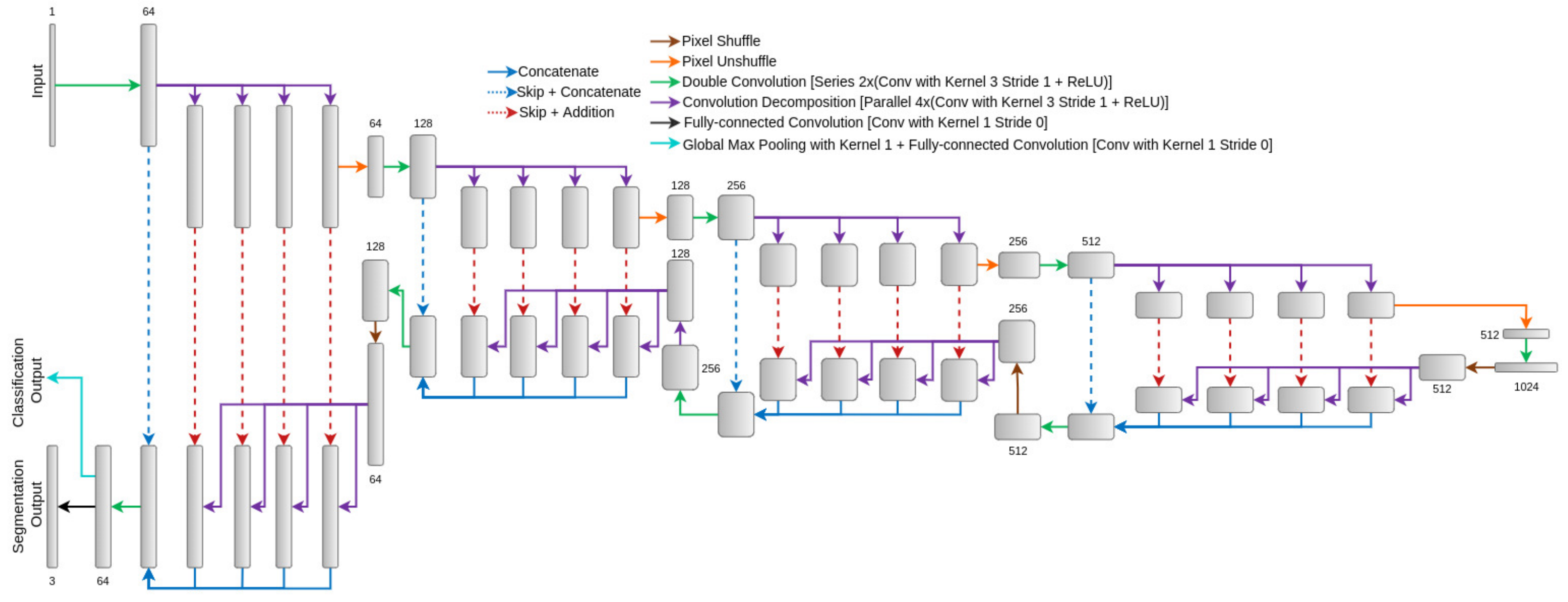}
\caption{The Network Architecture of the proposed GP-ShuffleUNet model\citep{chatterjee2021shuffleunet}. The original model was altered by adding a global pooling layer before the fully connected convolution in the training stage transforming the model to a GP-Model, while also adding a dropout layer with a probability value of 0.5.}
\label{fig:GP-ShuffleUNet}
\end{figure*}

Three different segmentation/reconstruction models were explored in this research: UNet~\citep{Ronneberger.5182015} - which was also used by \citet{Dubost.5222017}, ReconResNet~\citep{chatterjee2021reconresnet}, and ShuffleUNet~\citep{chatterjee2021shuffleunet}, to construct GP-UNet, GP-ReconResNet, and GP-ShuffleUNet, respectively. The resultant network architectures are shown in Figures~\ref{fig:GP-UNet},~\ref{fig:Recon},~and\ref{fig:GP-ShuffleUNet}, respectively. Original U-Net and ShuffleUNet models were modified by adding a dropout layer with a probability of 0.5 in the GP-models to avoid overfitting, whereas probability of the dropout layer which is already implemented in ReconResNet was increased to 0.5. The proposed GP-models were compared against two non-GP models: InceptionV3~\citep{szegedy2016rethinking} and ResNeXt50~\citep{xie2017aggregated}, which were chosen by comparing the performance of various other models. Table~\ref{Tab:MainResults_para} shows the number of parameters and the approximate training time. In addition to the GP-models, MProtoNet was included as a baseline model, which is an inherently interpretable classification model (No post-hoc explanation method needed). MProtoNet~\citep{wei2024mprotonet} provides transparency by comparing test samples against a learnt set of prototypical representations, allowing the model's decision to be explained in terms of similarity to these prototypes. Unlike the proposed GP-models, MProtoNet does not employ a global pooling mechanism and is not designed to produce localisation heatmaps as part of its architecture. Nevertheless, attribution maps derived from MProtoNet were post-processed in the same manner as the GP-model heatmaps to enable a fair comparison in the weakly-supervised segmentation setting. 

\begin{table}
  \caption{Number of parameters and training time for each model.}

	\centering
    \begin{tabular}{l c l}
      \toprule
      Model Name &  No. of Parameters  & Training Time \\     
      \midrule                                           
      InceptionV3       & 21.8 M      & $\sim2$ days   \\
      ResNext50         & 23.0 M      & $\sim2$ days   \\
      MProtoNet         & 8.83 M      & $\sim$ 4 days \\
      GP-UNet           & 1.90 M       & $\sim4$ days  \\
      GP-ShuffleUNet    & 26.4 M      & $\sim13$ days \\ %
      GP-ReconResNet    & 17.3 M      & $\sim9$ days  \\
      \toprule 
    \end{tabular}
    \label{Tab:MainResults_para}
\end{table}

\subsection{Implementation}  
The implementation was done using PyTorch and PyTorch-lightning, and the experiments were performed using an Nvidia GeForce RTX 2080 TI GPU. The data was partitioned into three sets: training, validation, and testing, with real-time spatial augmentation during the training phase. The networks were trained with the help of automatic mixed-precision for 300 epochs for the second dataset, with various batch sizes, depending on when each model converged and how large it was memory-wise, keeping the effective batch size as 128 by accumulating gradients of multiple batches before backpropagating. Specifically, the heavier GP-ShuffleUNet and GP-ReconResNet models utilised a physical batch size of 4 with 32 accumulation steps, GP-UNet used a batch size of 8 with 16 steps, and the baseline models (Inception V3, ResNeXt50, MProtoNet) employed a batch size of 32 with 4 steps. The loss was calculated using the weighted cross-entropy loss function and was optimised using the Adam optimiser with a learning rate of 0.001 and the betas were set to $(0.9, 0.999)$. To avoid overfitting, along with adding dropout layers inside the network models, L2 regularisation weights decay factor of 5e-4 was also employed. Finally, post-hoc interpretability methods~\citep{chatterjee2021torchesegeta} were applied to all GP and non-GP models - to be able to compare their interpretability. The code for this research is publicly available on GitHub \footnote{Code on GitHub: \url{https://github.com/soumickmj/GPModels}} and the trained model weights are available on Hugging Face\footnote{Weights on Hugging Face: \url{https://huggingface.co/collections/soumickmj/gp-models}}.

\subsection{Dataset}
\label{sec:dataset} 
The network models were employed for the task of brain tumour classification using two different datasets. The first dataset~\citep{JunCheng.2017} contains $3,064$ 2D T1-weighted contrast-enhanced MRI slices from 233 patients with three types of brain tumours: meningioma (708 slices), glioma (1426 slices), and pituitary tumour (930 slices), in three different orientations: axial (993 slices), coronal (1046 slices), and sagittal (1025  slices). The second dataset used in this research is the BraTS 2020 dataset~\citep{menze2014multimodal,lloyd2017high,bakas2018identifying}. This dataset contains multimodal scans acquired with different clinical protocols and scanners from 19 institutions of 346 patients with two different types of tumours: low-grade glioma (LGG, 73 patients) and high-grade glioma or glioblastoma (HGG, 273 patients), each having 3D volumes of T1-weighted (T1w), post-contrast T1-weighted (T1ce), T2-weighted (T2), and T2 Fluid Attenuated Inversion Recovery (T2-FLAIR, or simply, FLAIR) MRIs in axial orientation - a total of $4\times47,686$ slices after ignoring blank slices. The slices without any tumour were considered as "tumour-free" and a third class was created using them, resulting in $4,731$, $18,276$, and $24,679$ slices, for LGG, HGG, and tumour-free class, respectively. The images were manually segmented by one to four raters, and the annotations were agreed upon by neuro-radiology experts. All four types of MRIs were concatenated on the channel dimension and were supplied to the networks in 2D, slice-wise manner. The four-channel approach was inspired by the belief that different contrasts provide varying degrees of diagnosis strength. In a sense, the entire tumour or parts of it might appear clearer to the human eye and achieve greater evaluation results using one contrast versus the other\citep{Jeong.2014}. Three fold cross-validation was performed by randomly creating subsets with the ratio of 60:15:25 using stratified sampling for training, validation, and testing, respectively. Before feeding the images to the network, each of the images (or set of image contrasts in the case of BraTS) was normalised by dividing the values of that image (or image set) with its maximum intensity value.

\subsubsection{Augmentation}
Table~\ref{tab:margins} illustrates the real-time augmentation techniques performed only on the training set, in which each image in the set is randomly rotated in the range of -330 to +330, horizontally or vertically flipped, each with a 50\% chance of occurring independently from one another. Augmentation can be proven helpful to prevent overfitting and enhance model robustness \citep{krizhevsky2012imagenet,Wong.9282016}. 

\begin{table}
  \caption{The applied spatial augmentation techniques and their respective parameters.}
  \label{tab:margins}
	\centering
    \begin{tabular}{l l}
      \toprule
      Methods    & Parameter  \\
      \toprule                                           
      Flip horizontally        &   50\% probability \\
      Flip vertically          &   50\% probability \\
      Rotation             	   &   $ \pm 330^\circ$ \\
      \toprule 
    \end{tabular}
\end{table}


\subsection{Evaluation Criteria}
The F1-score was used as the primary quantitative metric for evaluating the models' classification performance. When having similar F1 scores, other measures such as confusion matrix, accuracy, recall, and precision were employed to quantify the results. To account for data imbalance in the BraTS dataset, a weighted average of the matrix values was used. The Dice similarity coefficient was used as a quantitative metric assessing the GP-models' segmentation performance. As for the last aspect, the model's inherent explainability and transparency were assessed by visually inspecting the generated localisation heatmaps, revealing the models' classification decision basis. Furthermore, post-hoc interpretability methods were applied to the best performing GP- and non-GP- models to be able to compare interpretability aspects in both.

\section{Results}

For the first set of experiments with the first dataset~\citep{JunCheng.2017}, trainings were performed by supplying individual 2D slices of T1ce MRIs as single-channel input to the network models and were trained to perform 3-class classification for the type of brain tumour: meningioma, glioma, and pituitary tumour. For the second set of experiments with the BraTS dataset~\citep{menze2014multimodal,lloyd2017high,bakas2018identifying}, trainings were performed by concatenating the four multi-modal MRIs: T1, T1ce, T2, and Flair on the channel-dimension, and were trained for 3-class classification for tumour-free and two different types of brain tumours - LGG and HGG. It is worth mentioning that both datasets are imbalanced, whereas the second dataset is much more imbalanced compared to the first one. The number of slices in the first dataset has a ratio of 23:30:47 for meningioma:pituitary:glioma classes, while the ratio in the second dataset was 10:38:52 for LGG:HGG:tumour-free classes. Even though the first dataset poses less challenge in terms of class imbalance, it provides two different challenges: it is considerably smaller and contains a mix of three orientations, while the second dataset is much larger and contain only MRIs acquired in axial orientation. The models were initially evaluated for their classification performance, followed by their segmentation performance. Finally, additional evaluations were performed with the interpretability techniques.

\subsection{Experiments with Dataset \#1}
\begin{figure*}           %
\centering
\includegraphics[width=0.9\textwidth]{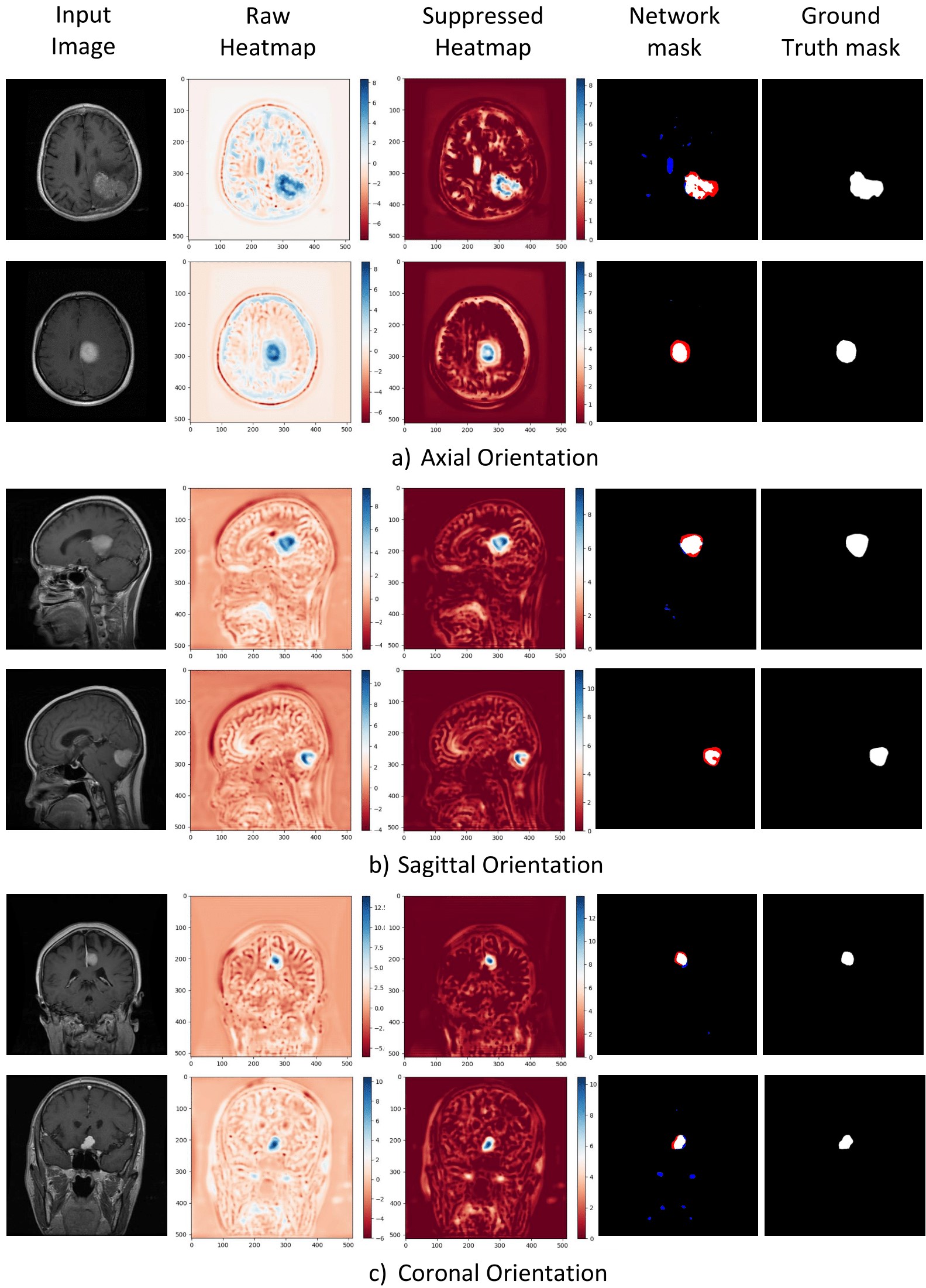}
\caption{Example results of the best-performing GP-ReconResNet on the 1st Dataset~\citep{JunCheng.2017}, in a) Axial, b) Sagittal and c) Coronal orientations. 1st column contains the input slices. 2nd column contains the model's prediction known as the raw heatmaps, where the red areas influenced the classification outcome negatively, and the blue areas influenced the classification outcome favourably. 3rd column contains the suppressed heatmaps, where negative values are suppressed to obtain positive attributions only. 4th column contains the network generated final masks = the suppressed heatmap + Otsu thresholding. The mask is then compared to the ground-truth mask, where white indicates true segmentation, blue indicates over-segmentation, and red indicates under-segmentation. 5th column contains the ground truth mask.}
\label{ExGPRecon}
\end{figure*}

Table~\ref{Tab:PreResults1} shows the quantitative results for the different GP-models and the non-GP baseline models. The GP-ReconResNet performed the best among the GP-models, scoring an F1-score of $0.95$, while the GP-UNet and GP-ShuffleUNet secured the second and third positions, respectively, with $0.85$ and $0.82$ F1-scores. Fig.~\ref{ExGPRecon} shows a few examples of the segmentation in all three orientations for the GP-ReconResNet. It can be observed that the heatmap shows the tumour appropriately localised, and the post-processing helped in obtaining the segmentations. 

\begin{table}
  \caption{Resulting classification metrics (Precision, Recall, F1-score, Accuracy) for all the models, and the segmentation metrics (median Dice score) for the GP-models, on the first dataset~\citep{JunCheng.2017}}
	\centering
	\resizebox{0.48\textwidth}{!}{
	\begin{tabular}{l c c c c c}
      \toprule
      Model Name & Precision  & Recall  & F1-Score & Accuracy & Med (DSC)\\ 
      \midrule                                           
      InceptionV3   & 0.96   & 0.97  & 0.97   & 0.97 & -\\     
      ResNext50     & 0.96   & 0.96  & 0.96   & 0.97 & - \\    
      GP-ReconResNet & 0.96   & 0.95  & 0.95   & 0.95 & $ 0.10 \pm 0.0365$ \\ 
      GP-ShuffleUNet & 0.94  & 0.92  & 0.93  & 0.93     & 0.04 $\pm$ 0.0069  \\
      GP-UNet        & 0.86   & 0.85  & 0.85   & 0.85 & $ 0.05 \pm 0.0115$ \\ 
      \toprule 
    \end{tabular}}
    \label{Tab:PreResults1}
\end{table}

\subsection{Experiments with Dataset \#2: BraTS 2020 dataset}
On the BraTS dataset, the models show a different trend compared to the first dataset. GP-UNet and GP-ShuffleUNet resulted in $0.93$ F1-score, while GP-ReconResNet resulted in $0.92$. In terms of accuracy, GP-ShuffleUNet got $0.94$, while GP-UNet and GP-ReconResNet both obtained $0.93$. However, both the non-GP baselines resulted in $0.95$ F1-score and accuracy. Even though the non-GP models resulted in better scores, it is worth mentioning that the non-GP models are not inherently explainable and do not assist in producing segmentations as the GP-models. Table~\ref{Tab:MainResults3} shows the complete results on different metrics. 

\begin{figure*}[hbtp]           %
\centering
\includegraphics[width=0.8\textwidth]{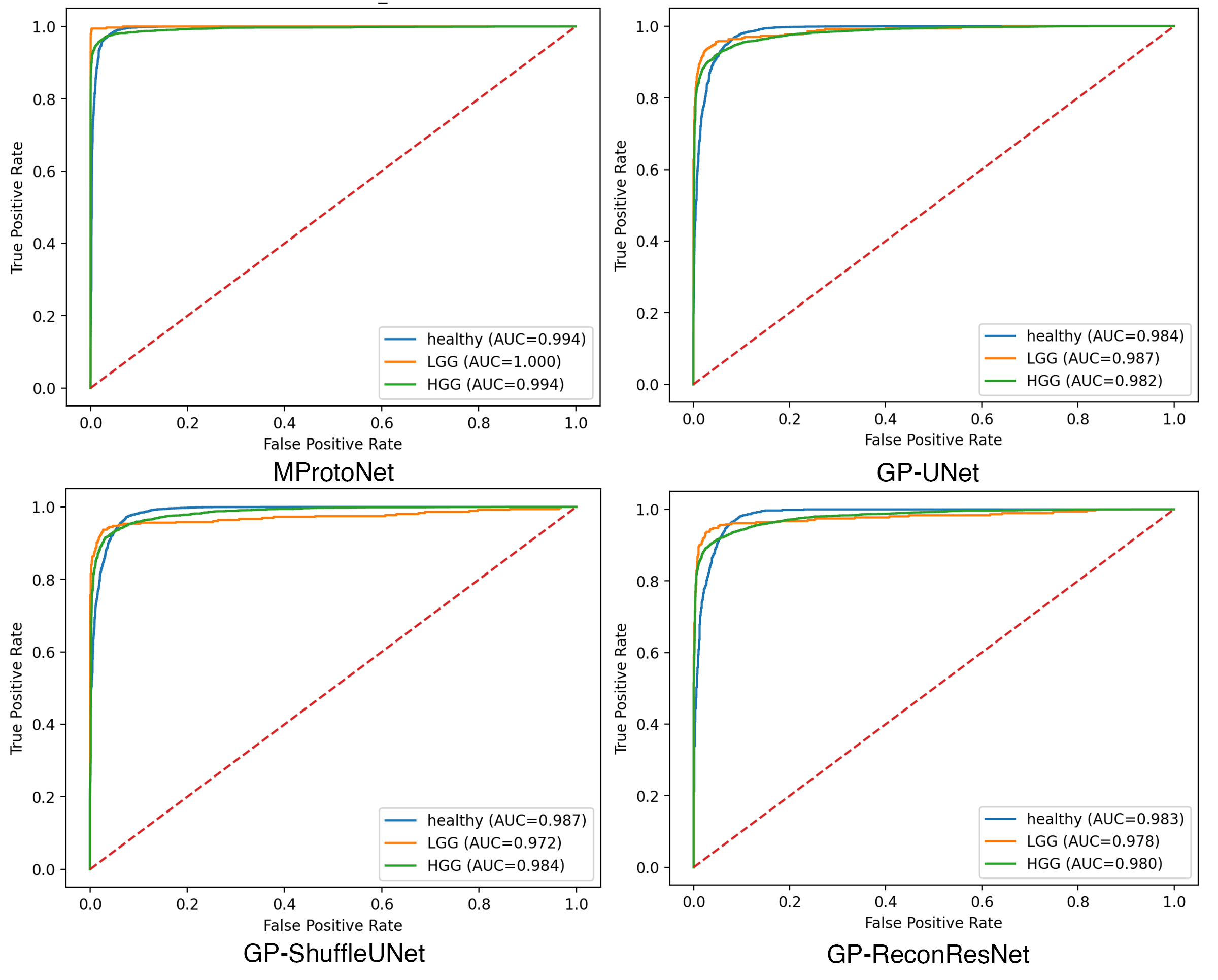}
\caption{Receiver Operating Characteristic (ROC) curves indicating global separability of tumour subclasses. The plots depict the diagnostic discrimination of the inherently explainable models. While near-perfect AUC values suggest strong generalisation, they must be interpreted with caution due to the insensitivity of ROC metrics to class imbalance.}
\label{fig:AUC}
\end{figure*}

\begin{figure*}[hbtp]           %
\centering
\includegraphics[width=0.8\textwidth]{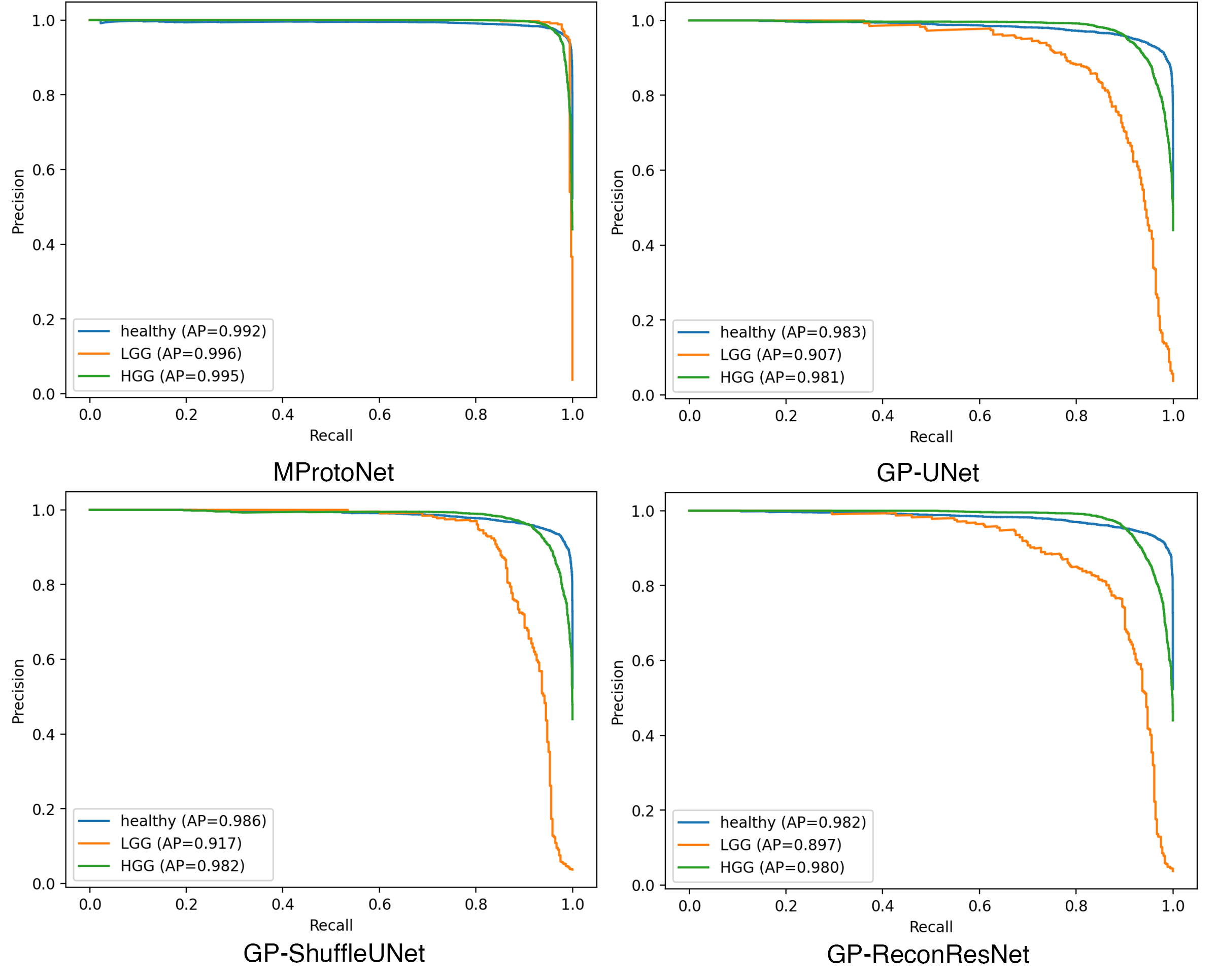}
\caption{Precision-Recall (PR) analysis demonstrating diagnostic robustness. Unlike the ROC curves, these plots expose the performance trade-offs on the minorityLGG class. The variance in LGG precision (e.g. GP-ReconResNet AP=0.897) highlights the specific challenge of detecting under-represented pathologies, providing a rigorous validation of the models' confidence thresholds.}
\label{fig:PR}
\end{figure*}

To provide a more rigorous assessment of diagnostic robustness beyond scalar metrics, the decision boundaries using Receiver Operating Characteristic (ROC) and Precision-Recall (PR) curves were analysed. Figure \ref{fig:AUC} illustrates the ROC curves, where the Area Under the Curve (AUC) approaches unity for most architectures, suggesting excellent global separability between healthy tissue, LGG, and HGG classes. Crucially, DeLong's test revealed no statistically significant difference ($p > 0.05$) between the classification performance of any of these models, confirming that the proposed GP-models achieve diagnostic separability comparable to the baseline architectures despite their added explainability constraints. However, given the significant class imbalance inherent to the BraTS dataset (where LGG represents the minority class), ROC metrics can present an overly optimistic view of performance due to the large volume of true negatives. Consequently, Figure \ref{fig:PR} presents the Precision-Recall (PR) curves, which offer a more conservative and clinically relevant evaluation. Whilst the models maintain high fidelity for the majority classes, the PR analysis unmasks the latent difficulty in characterising the LGG subtype. Notably, the drop in Average Precision (AP) for LGG, particularly visible in the GP-ReconResNet architecture (AP=0.897), visually quantifies the challenge of minority class detection that is otherwise obscured in the tabular results (Table \ref{Tab:MainResults3}). This distinction is critical for weakly-supervised segmentation, as the reliability of the generated heatmaps is intrinsically linked to the precision of these classification confidence thresholds.

\begin{figure*}           %
\centering
\includegraphics[width=\textwidth]{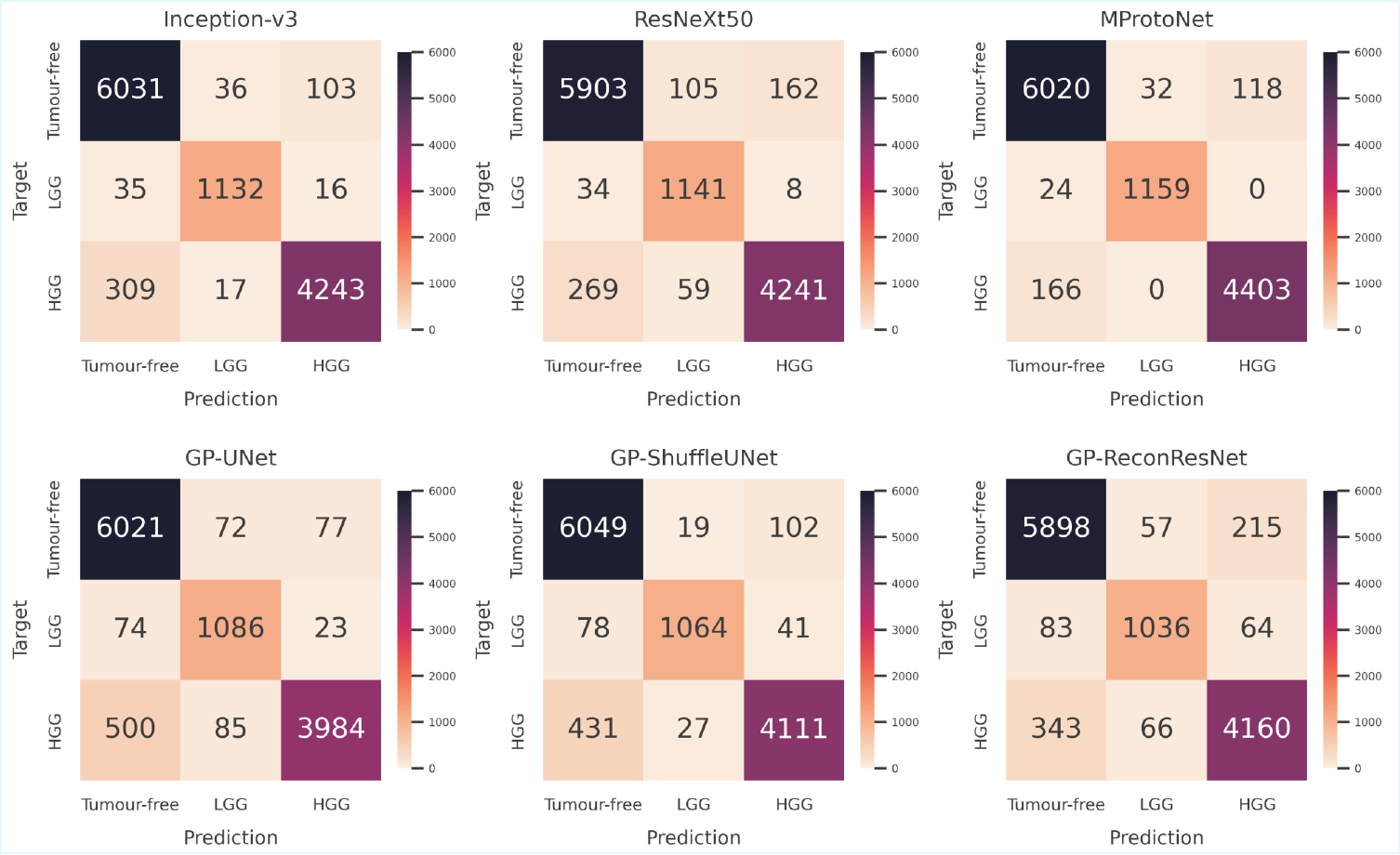}
\caption{Confusion matrices of the GP- and non-GP- models, where the three classification labels depicted as HGG for high-grade glioma (glioblastoma), LGG for low-grade glioma, and tumour-free.}
\label{fig:CM}
\end{figure*}

The classification performance of these models was further shown using confusion matrices in Fig.~\ref{fig:CM}. The accuracy on a subset of the test set considering only the slices for the two tumour classes for the GP-models were compared against other state-of-the-art methods which also have used the BraTS dataset (but different versions), and it can be seen that the GP-models outperform all the other methods, while also being inherently explainable - shown in Table~\ref{Tab:MainResults5}. The reason behind considering only LGG and HGG classes in this analysis was that the compared methods only have considered these classes in their analyses. It is to be noted, though, that the methods that worked with 3D volumes included the tumour-free slices - part of the volumes but classified the whole volume as LGG or HGG class.

\begin{table}
  \caption{Resulting classification metrics (Precision, Recall, F1-score, Accuracy) for all the models, and the segmentation metrics [Median (95\% CI) Dice score] for the GP-models, on the BraTS 2020 dataset.}

	\centering
  \resizebox{0.48\textwidth}{!}{%
  \begin{tabular}{@{}lccccc@{}}
\toprule
Model Name        & Precision    & Recall       & F1-Score     & Accuracy     & Dice score       \\ \midrule
InceptionV3       & 0.95  & 0.95  & 0.95  & 0.95  & -               \\
ResNeXt50         & 0.95  & 0.95  & 0.95  & 0.95  & -               \\
MProtoNet         & 0.97  & 0.97  & 0.97  & 0.97  & 0.121 (0.118--0.125) \\
GP-UNet           & 0.93  & 0.93  & 0.93  & 0.93  & 0.728 (0.715--0.739) \\
GP-ShuffleUNet    & 0.94  & 0.94  & 0.93  & 0.94  & 0.720 (0.707--0.731) \\
GP-ReconResNet$_1$ & 0.93  & 0.93  & 0.92  & 0.93  & 0.647 (0.638--0.658) \\
GP-ReconResNet$_2$ & 0.93  & 0.93  & 0.92  & 0.93  & 0.689 (0.677--0.700) \\ \bottomrule
\end{tabular}}
    \label{Tab:MainResults3}
\end{table}

\begin{table}
  \caption{Comparison of the proposed GP-models with the previously published non-deep learning (non-DL) and deep learning (DL) works ($\dagger$ cross-validated) for the BraTS dataset.}

	\centering
	\resizebox{0.48\textwidth}{!}{%
    \begin{tabular}{l l c c c}
      \toprule
      Study / Model &	Contrast  &	\begin{tabular}[c]{@{}c@{}}Model \\ Type\end{tabular} & Explainable  & 	\begin{tabular}[c]{@{}c@{}}Test \\ Accuracy\end{tabular}\\     
      \midrule                                           
      \citet{latif2017multiclass}   & \begin{tabular}[c]{@{}l@{}}T1, T1ce, T2, \\ T2-FLAIR\end{tabular}                   & \begin{tabular}[c]{@{}c@{}}non-DL \\ 3D\end{tabular}  &  N  & 88.31 \% \\
      \citet{cho2017classification}   & \begin{tabular}[c]{@{}l@{}}T1, T1ce, T2, \\ T2-FLAIR\end{tabular}                   & \begin{tabular}[c]{@{}c@{}}non-DL \\ 3D\end{tabular}  &  N  & $89.81 \%^{\dagger}$ \\
      \citet{shahzadi2018cnn}   & T2-FLAIR                   & DL 3D  &  N  & 84.00 \% \\
      \citet{ge2018deep}         & \begin{tabular}[c]{@{}l@{}}T1, T2 , \\ T2-FLAIR\end{tabular}          & DL 2D  &  N  & 90.87\%\\
      \citet{yang2018glioma}       & T1ce                       & DL 2D  &  N  & $94.50\%^{\dagger}$\\
      \citet{mzoughi2020deep}   & T1ce                       & DL 3D  &  N  & 96.49\%\\
      \citet{zhuge2020automated}     & \begin{tabular}[c]{@{}l@{}}T1, T1ce, T2, \\ T2-FLAIR\end{tabular}     & DL 3D  &  N  & 97.10\% \\ 
      \citet{chatterjee2021spatiotemp}  & T1ce  & DL 3D  &  N  & $96.98\%^{\dagger}$ \\
      \citet{barstugan2023classification}  & \begin{tabular}[c]{@{}l@{}}T1, T1ce , \\ T2\end{tabular}  & \begin{tabular}[c]{@{}c@{}}non-DL \\ 3D\end{tabular}  &  Y  & $90.17\%^{\dagger}$ \\
      \citet{hafeez2023cnn}     & \begin{tabular}[c]{@{}l@{}}T1, T1ce, T2, \\ T2-FLAIR\end{tabular}     & DL 2D  &  N  & $97.15\%$ \\ 
      \citet{wei2024mprotonet}     & \begin{tabular}[c]{@{}l@{}}T1, T1ce, T2, \\ T2-FLAIR\end{tabular}     & DL 2D  &  N  & $86.80\%^{\dagger}$ \\ 
      \citet{dutta2024arm}     & \begin{tabular}[c]{@{}l@{}}T1, T1ce, T2, \\ T2-FLAIR\end{tabular}     & DL 2D  &  N  & $96.87\%^{\dagger}$ \\ 
    MProtoNet \citep{wei2024mprotonet}    & \begin{tabular}[c]{@{}l@{}}T1, T1ce, T2, \\ T2-FLAIR\end{tabular}     & DL 2D  &  Y  & $97.79\%$\\
      \textbf{GP-UNet}           &\textbf{\begin{tabular}[c]{@{}l@{}}T1, T1ce, T2, \\ T2-FLAIR\end{tabular}}     &\textbf{DL 2D}  &  \textbf{Y}  & \textbf{$98.27\%^{\dagger}$}\\
     \textbf{GP-ShuffleUNet}    &\textbf{\begin{tabular}[c]{@{}l@{}}T1, T1ce, T2, \\ T2-FLAIR\end{tabular}}    & \textbf{DL 2D}  &  \textbf{Y}  & \textbf{$98.74\%^{\dagger}$}\\
     \textbf{GP-ReconResNet}    & \textbf{\begin{tabular}[c]{@{}l@{}}T1, T1ce, T2, \\ T2-FLAIR\end{tabular}}     & \textbf{DL 2D}  &  \textbf{Y}  & \textbf{$97.99\%^{\dagger}$}\\
      \toprule 
    \end{tabular}}
    \label{Tab:MainResults5}
\end{table}

\begin{figure*}[htbp]
    \centering
    
    \begin{subfigure}[b]{0.48\textwidth}
        \centering
        \includegraphics[width=\linewidth]{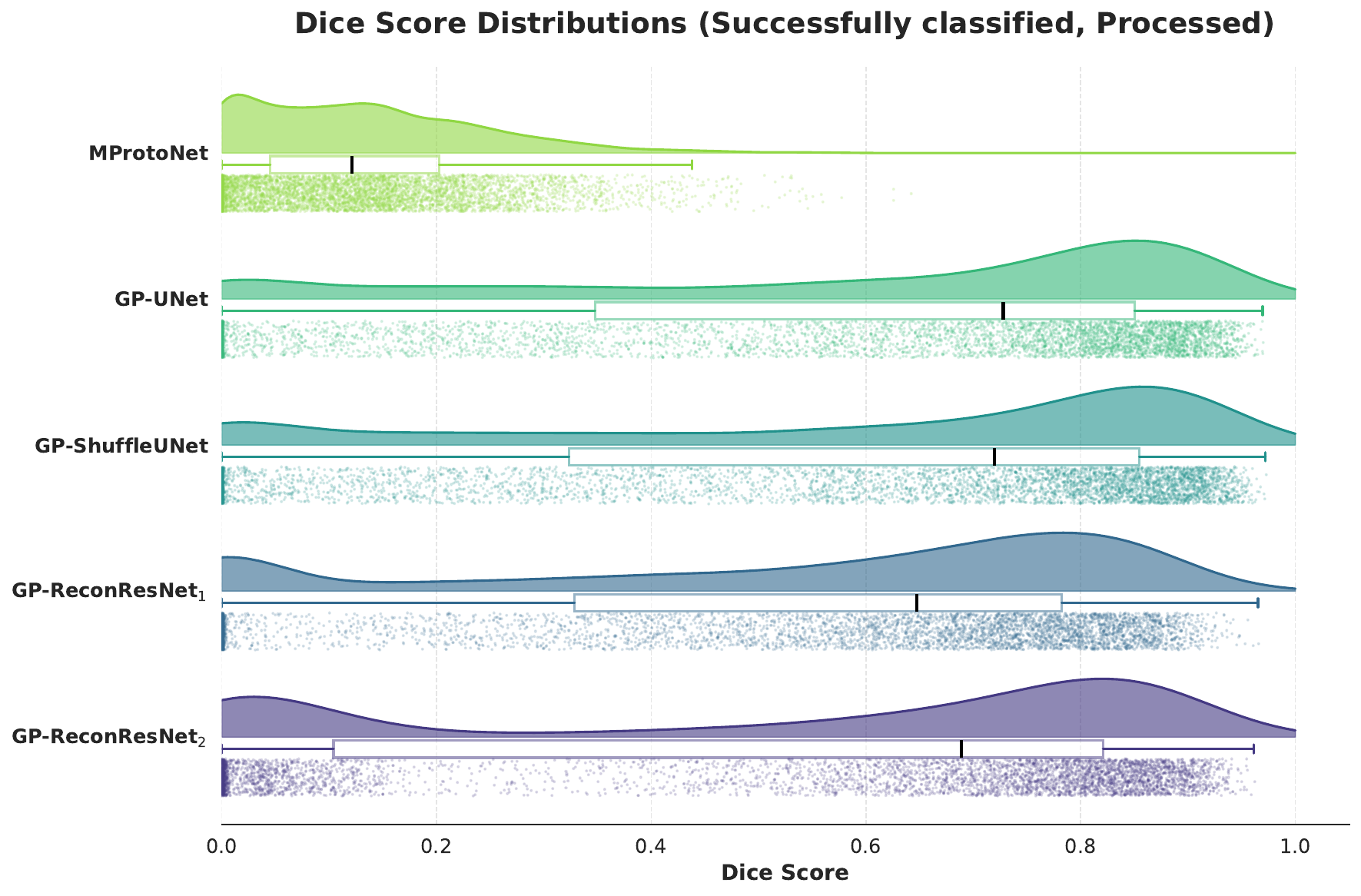}
        \caption{} 
        \label{fig:sub1}
    \end{subfigure}    
    \hfill     
    \begin{subfigure}[b]{0.48\textwidth}
        \centering
        \includegraphics[width=\linewidth]{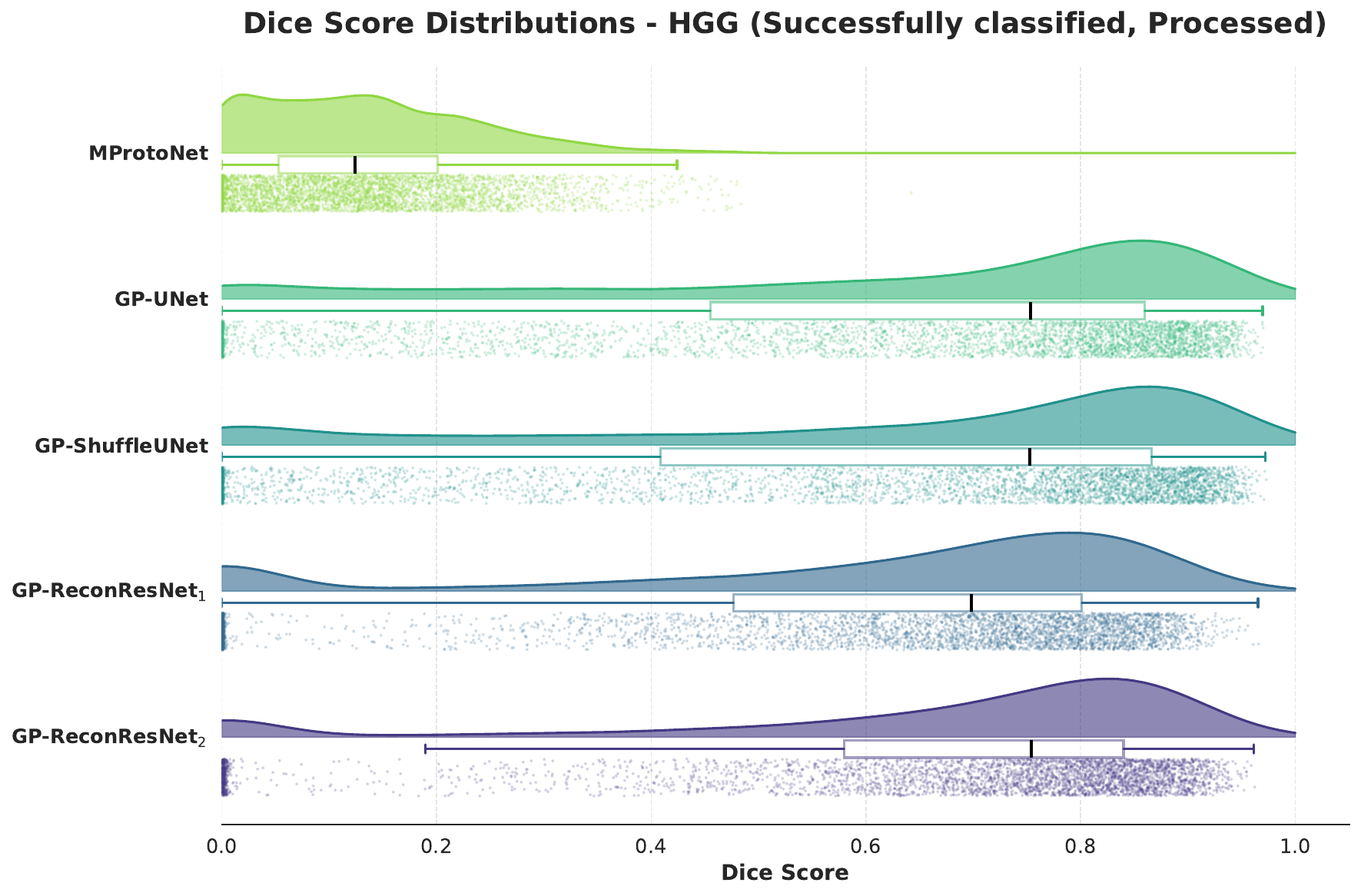}
        \caption{} 
        \label{fig:sub2}
    \end{subfigure}
    
    \caption{Comparative distribution of segmentation fidelity for the proposed Gaussian Process architectures versus the baseline. \textbf{(a)}: Raincloud plots displaying the distribution of post-processed Dice scores for the subset of successfully classified images across the full dataset. The plots comprise a half-violin estimating the probability density, a box plot indicating the median and interquartile range, and jittered points representing individual slice scores. All four GP-based models exhibit a statistically significant distributional shift towards higher accuracy compared to the MProtoNet baseline ($p < 0.001$), with GP-UNet showing the highest overall density. \textbf{(b)} Stratified analysis focusing exclusively on the High-Grade Glioma (HGG) subtype. In this predominant pathological class, the reconstruction-based GP-ReconResNet$_2$ demonstrates superior performance, evinced by a tighter density concentration near the upper bound of the metric, significantly outperforming its predecessor GP-ReconResNet$_1$ and surpassing the U-Net variants.}
    \label{fig:rainclouds}
\end{figure*}

\begin{figure}           
\centering
\includegraphics[width=0.49\textwidth]{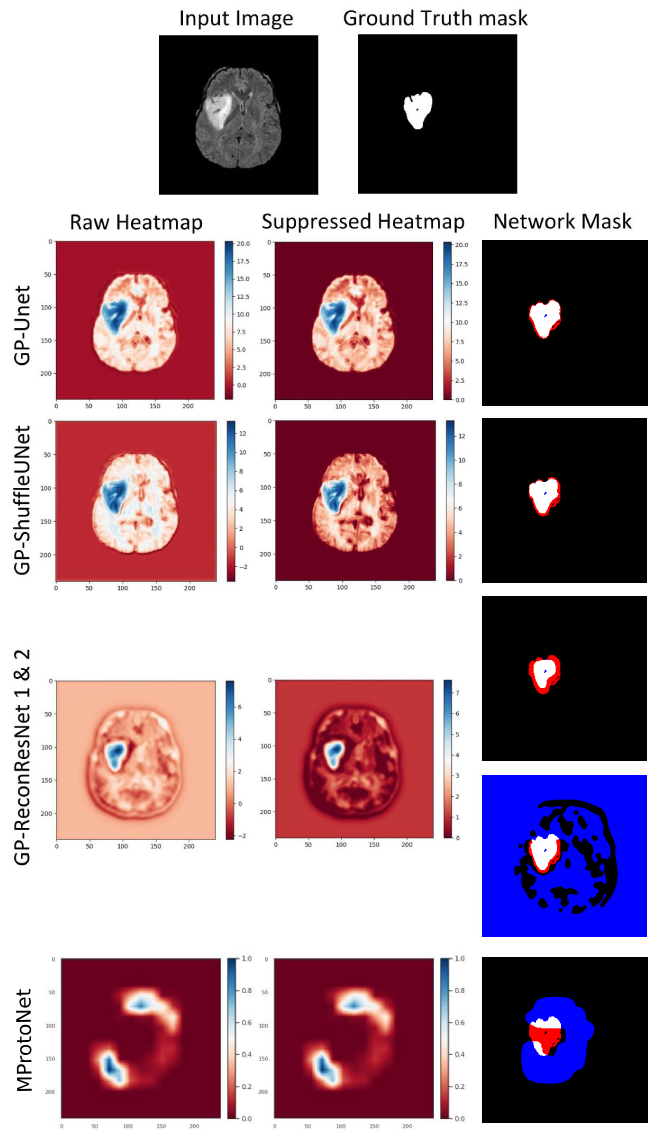}
\caption{Example results of the interpretable models' (GP-models and baseline MProtoNet) correctly classified LGG tumour (class 1) from the BraTS 2020 dataset. 1st row is the input slice, the ground truth mask is used for comparison, and the rest of the rows are the models' outputs. For the rest of the rows: 1st column contains the models' predictions known as the raw heatmaps, where the red areas influenced the classification outcome negatively, and the blue areas influenced the classification outcome favourably; 2nd column contains the suppressed heatmaps, where negative values are suppressed to obtain positive attributions only; 3rd column contains the networks' generated final masks = the suppressed heatmap + thresholding.}
\label{fig:MainResults2}
\end{figure}

\begin{figure}          
\centering
\includegraphics[width=0.49\textwidth]{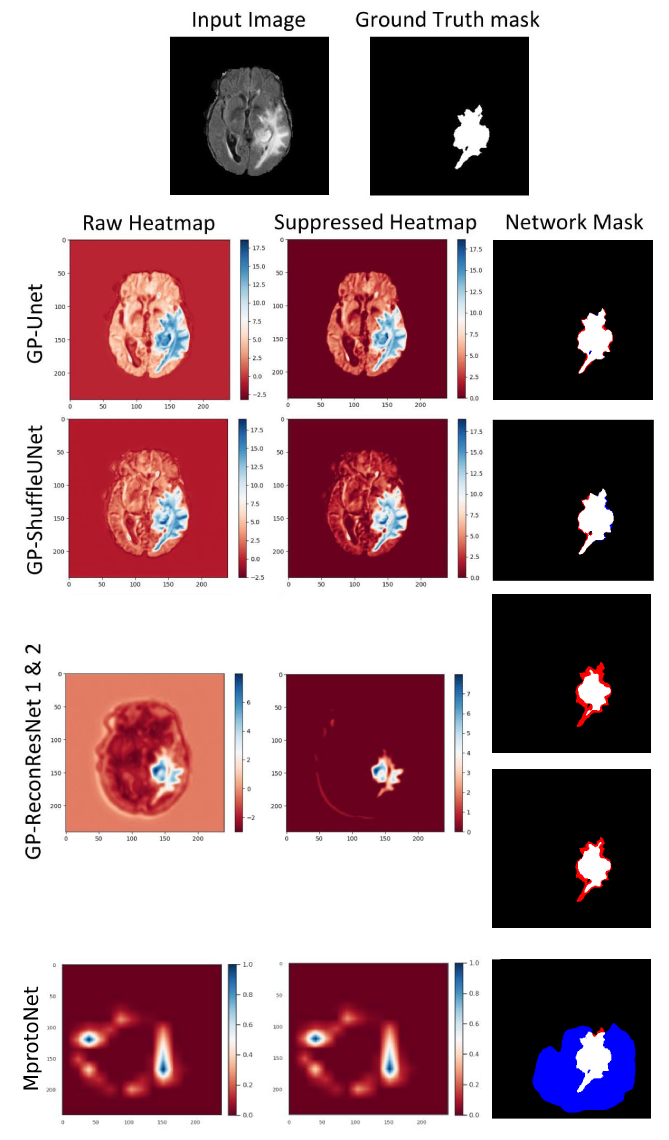}
\caption{Example results of the interpretable models' (GP-models and baseline MProtoNet) correctly classified HGG tumour (class 2) from the BraTS 2020 dataset. 1st row is the input slice, the ground truth mask is used for comparison, and the rest of the rows are the models' outputs. For the rest of the rows: 1st column contains the models' predictions known as the raw heatmaps, where the red areas influenced the classification outcome negatively, and the blue areas influenced the classification outcome favourably; 2nd column contains the suppressed heatmaps, where negative values are suppressed to obtain positive attributions only; 3rd column contains the networks' generated final masks = the suppressed heatmap + thresholding.}
\label{fig:MainResults4}
\end{figure}

Regarding segmentation performance, the Dice scores exhibit a divergent trend compared to classification metrics (Table~\ref{Tab:MainResults3}). To generate binary masks from the raw heatmaps, a dynamic multi-level Otsu thresholding technique using three classes was employed, selecting the highest-intensity threshold to isolate the tumour. This threshold was fine-tuned via a scalar offset ($Threshold - Offset$) optimised on the validation set before applying the same threshold to the final test set, followed by lightweight post-processing (morphological closing, hole filling, and component filtering). This method was compared against a fixed threshold baseline of 0.5, followed by the same post-processing scheme. Whilst the dynamic strategy favoured GP-UNet and GP-ShuffleUNet, the fixed threshold proved optimal for GP-ReconResNet; hence, results for the latter are reported for both approaches (GP-ReconResNet$_1$ and GP-ReconResNet$_2$). 
The resulting distributions of the Dice scores have been visualised using raincloud plots in Fig.~\ref{fig:rainclouds}. Figures~\ref{fig:MainResults2}~and~\ref{fig:MainResults4} show the generated heatmaps and the segmentation results of the GP-models, for LGG and HGG classes, respectively.

  \begin{table*}[]
  \caption{Segmentation performance of the proposed model variants. Results are compared across the full dataset (All) and the subset of successfully classified images, both before (Raw) and after post-processing the predicted segmentation masks. Values are reported as Median (95\% CI).}
  \centering
\begin{tabular}{@{}lcccc@{}}
\toprule
                   & \multicolumn{2}{c}{All}                   & \multicolumn{2}{c}{Successfully classified} \\ \cmidrule(l){2-5}
                   & Raw                 & Processed           & Raw                  & Processed            \\ \midrule
GP-UNet             & \textbf{0.598 (0.581–0.612)}     & 0.669 (0.654–0.687)     & \textbf{0.656 (0.642–0.671)}      & \textbf{0.728 (0.715–0.739) }     \\
GP-ShuffleUNet      & 0.593 (0.577–0.607)     & \textbf{0.671 (0.656–0.686)}     & 0.645 (0.632–0.656)      & 0.720 (0.707–0.731)      \\
GP-ReconResNet$_1$  & 0.564 (0.554–0.576)     & 0.607 (0.595–0.618)     & 0.606 (0.596–0.615)      & 0.647 (0.638–0.658)      \\
GP-ReconResNet$_2$  & 0.567 (0.553–0.579)     & 0.641 (0.628–0.656)     & 0.613 (0.602–0.626)      & 0.689 (0.677–0.700)      \\
\bottomrule
\end{tabular}
\label{Tab:GP_Dice_FullDS}
\end{table*}

\begin{table*}[]
  \caption{Segmentation performance of the proposed model variants only for the HGG class. Results are compared across the full dataset (All) and the subset of successfully classified images, both before (Raw) and after post-processing the predicted segmentation masks. Values are reported as Median (95\% CI).}
  \centering
\begin{tabular}{@{}lcccc@{}}
\toprule
                   & \multicolumn{2}{c}{All}                   & \multicolumn{2}{c}{Successfully classified} \\ \cmidrule(l){2-5}
                   & Raw                 & Processed           & Raw                  & Processed            \\ \midrule
GP-UNet             & 0.630 (0.612–0.645)     & 0.697 (0.684–0.712)     & 0.689 (0.679–0.703)      & 0.753 (0.742–0.764)      \\
GP-ShuffleUNet      & 0.639 (0.624–0.653)     & 0.707 (0.692–0.723)     & 0.678 (0.669–0.691)      & 0.753 (0.742–0.764)      \\
GP-ReconResNet$_1$  & 0.637 (0.629–0.645)     & 0.673 (0.660–0.682)     & 0.665 (0.657–0.671)      & 0.698 (0.690–0.707)      \\
GP-ReconResNet$_2$  & \textbf{0.662 (0.654–0.671)}     & \textbf{0.733 (0.724–0.741)}     & \textbf{0.689 (0.681–0.696)}      & \textbf{0.754 (0.749–0.760)}      \\
\bottomrule
\end{tabular}
\label{Tab:GP_Dice_HGG}
\end{table*}


\subsubsection{Statistical evaluation of segmentation performance}
To rigorously assess the segmentation fidelity and statistical significance of the proposed architectures, a non-parametric bootstrap approach ($n=10,000$ iterations) was employed to estimate the $95\%$ confidence intervals (CI) of the median Dice scores, alongside the Wilcoxon signed-rank test for paired samples. As presented in Table \ref{Tab:MainResults3}, which details the results for successfully classified images following post-processing, all four GP models: GP-UNet, GP-ShuffleUNet, GP-ReconResNet$_1$, and GP-ReconResNet$_2$, demonstrated a substantial and statistically significant improvement over the baseline MProtoNet ($p < 0.001$; bootstrap CIs strictly excluding zero). Specifically, the proposed models exhibited median improvements ranging from approximately $0.54$ to $0.61$ in this scenario, validating the superior segmentation capability of the GP-based approach over the prototype-based baseline. While the subsequent tables (Tables \ref{Tab:GP_Dice_FullDS} and \ref{Tab:GP_Dice_HGG}) focus exclusively on the comparative performance of the GP model variants, it is important to note that the statistical evaluations confirmed this significant advantage against MProtoNet across all tested scenarios.

Comparisons amongst the proposed architectures on the global dataset, summarized in Table \ref{Tab:GP_Dice_FullDS}, revealed that the U-Net-based variants consistently outperformed the ResNet-based models. GP-UNet achieved the highest median Dice scores, recording 0.598 ($95\%$ CI: $0.581$-$0.612$) for raw predictions on the full dataset and 0.669 ($95\%$ CI: $0.654$-$0.687$) after post-processing. This was closely followed by GP-ShuffleUNet (raw: 0.593), where statistical tests indicated no significant difference in the raw output (median difference $0.005$; bootstrap CI $[-0.008, 0.017]$). Conversely, the ResNet-based formulations generally trailed the U-Net variants on the global metrics. However, within the ResNet pairing, GP-ReconResNet$_2$ demonstrated a distinct advantage over GP-ReconResNet$_1$ (0.564 vs 0.567 raw), particularly in processed metrics where the improvement was statistically significant ($p < 0.001$).

Given the significant class imbalance within the dataset, where HGG accounts for $38.3\%$ of the slices compared to just $9.9\%$ for LGG, a stratified analysis was conducted on the HGG subtype to evaluate model performance on the predominant pathological presentation, presented in Table \ref{Tab:GP_Dice_HGG}. This stratification isolates the architectural efficacy on the sufficiently represented class, mitigating the high variance associated with the sparse LGG samples. Consistent with the aggregated metrics, all four proposed models maintained a substantial lead over MProtoNet, yielding median improvements in raw Dice scores between $0.51$ and $0.54$ ($p < 0.001$, Wilcoxon signed-rank test; bootstrap CIs excluding zero). However, unlike the overall dataset where U-Net variants often prevailed, the HGG subset favoured the ReconResNet architecture. Specifically, GP-ReconResNet$_2$ emerged as the most robust model for the majority pathology class, significantly outperforming its GP-ReconResNet$_1$ with a median difference of $0.025$ in raw scores and a marked $0.060$ increase in processed scores ($p < 0.001$), achieving a raw median Dice score of 0.662 ($95\%$ CI: $0.654$-$0.671$) and a processed score of 0.733 ($95\%$ CI: $0.724$-$0.741$). Furthermore, GP-ReconResNet$_2$ surpassed the otherwise competitive GP-UNet by a median margin of $0.032$ (raw) and $0.036$ (processed). Meanwhile, the performance difference between the two U-Net variants remained statistically negligible under bootstrap estimation (median difference $< 0.01$), reinforcing the finding that while the GP component is critical for improvement over the baseline, the ReconResNet offers specific benefits for delineating the complex structures of the primary tumour subtype.

\subsection{Post-hoc interpretability}
\begin{figure}          
\centering
\includegraphics[width=0.5\textwidth]{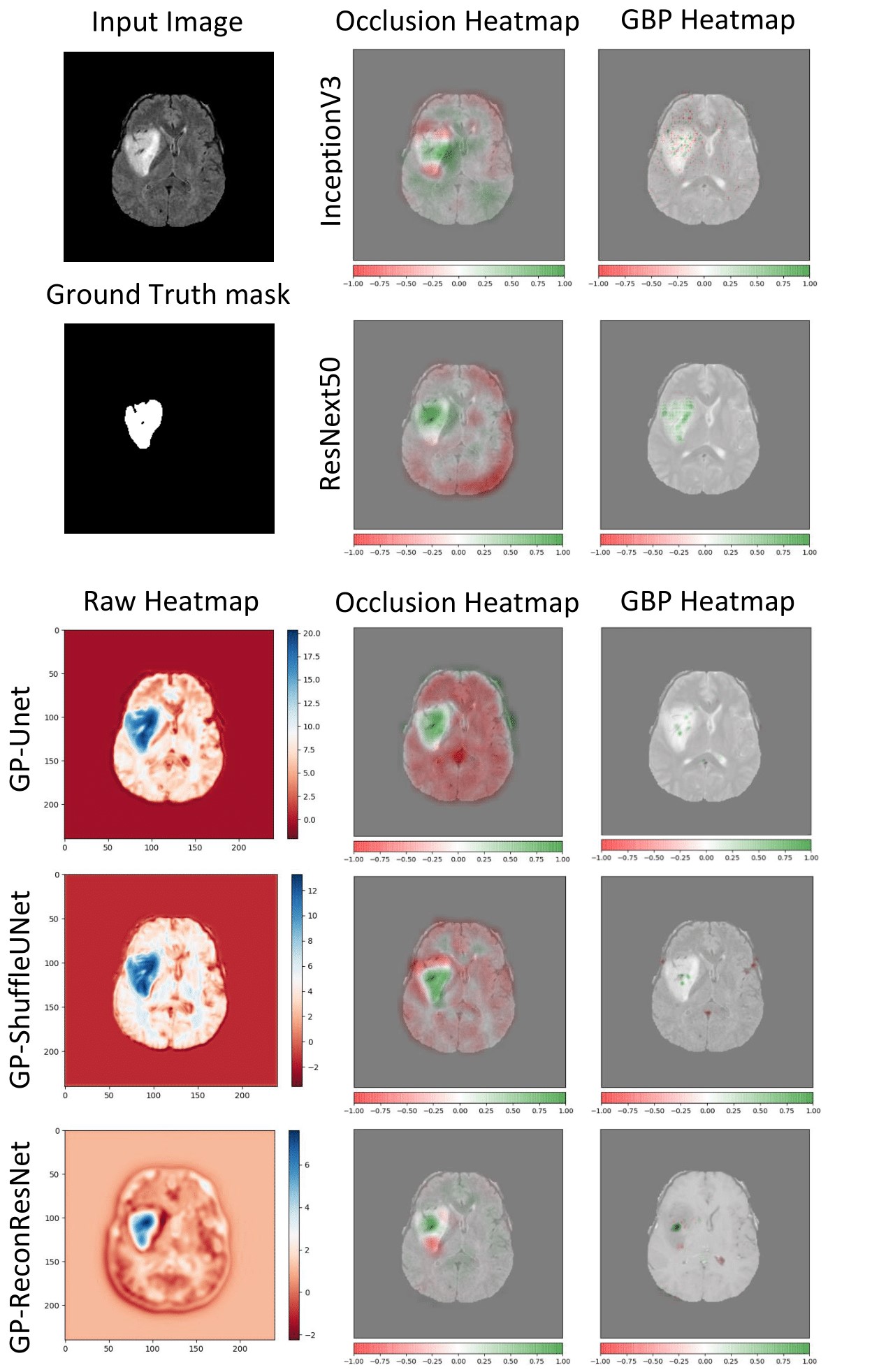}
\caption{Example interpretability results of the GP- and non-GP models, correctly classified LGG (class 1). 1st column contains the input image, ground-truth mask, and the raw heatmaps of the GP-models. 2nd and 3rd columns contain the post-hoc interpretability attributions using the occlusion and the guided backpropagation (GBP) methods, respectively - overlaid as heatmaps on the input slices.}
\label{fig:Inter_2}
\end{figure}

\begin{figure}           
\centering
\includegraphics[width=0.5\textwidth]{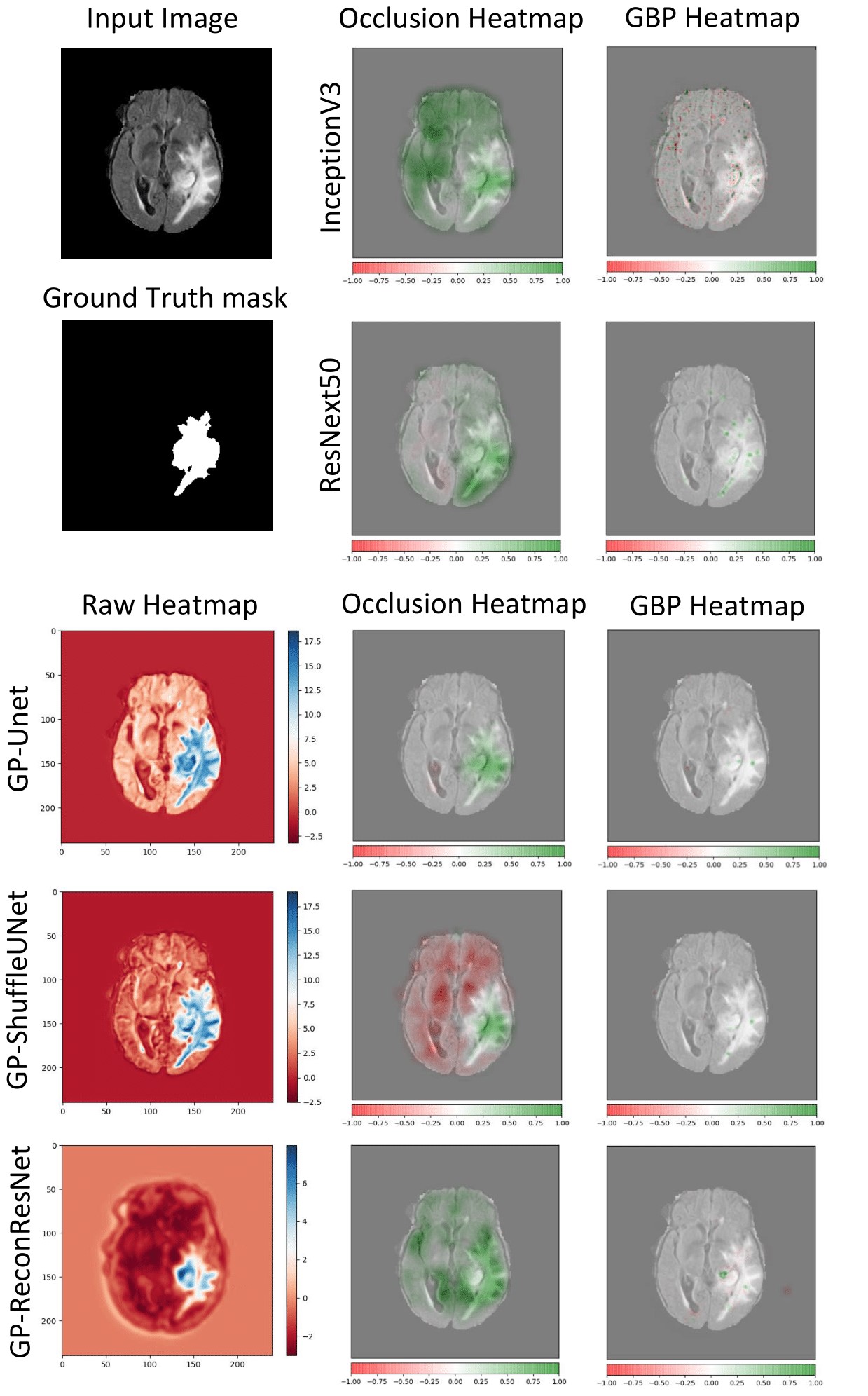}
\caption{Example interpretability results of the GP- and non-GP models, correctly classified HGG (class 2). 1st column contains the input image, ground-truth mask, and the raw heatmaps of the GP-models. 2nd and 3rd columns contain the post-hoc interpretability attributions using the occlusion and the guided backpropagation (GBP) methods, respectively - overlaid as heatmaps on the input slices.}
\label{fig:Inter_3}
\end{figure}

The GP-models presented here are explainable by nature, giving them an advantage over the competition in terms of model transparency and explainability portrayed. The focus area of the network can be understood from the generated heatmaps by those GP-models. The non-GP baselines are not explainable or interpretable, and typically post-hoc interpretability methods can be applied to understand the focus area of those networks~\citep{chatterjee2020exploration}. Therefore, two interpretability methods, one feature-based model attribution technique - occlusion, and one gradient-based model attribution technique - guided backpropagation, were applied to all the models to generate the interpretability results using the TorchEsegeta pipeline~\citep{chatterjee2021torchesegeta}. Figures~\ref{fig:Inter_2}~and~\ref{fig:Inter_3} show the interpretability results for all the five models, for LGG and HGG, respectively. As the GP-models are inherently explainable, the resultant heatmaps are also compared against the interpretability results. 

For the GP-models, by comparing the heatmaps against the interpretability results, it can be said that the interpretability results do agree with the heatmaps. The occlusion results show that the GP-models focused on the tumour. For GP-ReconResNet however, some questionable results can be seen. For LGG class, some negative attributions can be seen in the bottom side of the tumour, while for HGG class, the focus is scattered in different parts of the brain - but missing parts of the tumour, which is shown as the most important focus region according to the heatmap. Guided backpropagation has also shown the centre of the focus area(s) of the network, and they agree with the heatmaps. However, the area covering the attribution is considerably smaller than occlusion, as well as the heatmaps. This analysis shows that the interpretability results can be trusted, but might be considered with caution. 

The interpretability results of the non-GP baselines show that the reasoning done by the ResNeXt50 model is better than the InceptionV3, even though they both have secured the same classification scores. The positive, as well as the negative attributions for the InceptionV3, are scattered nearly all over the brain.

\section{Discussion}
\label{ch:discussion}

\noindent The principal objective of this study was to bridge the dichotomy between high-performance "black-box" diagnostics and clinically trustworthy, transparent systems. The proposed GP-models have demonstrated classification performance comparable to non-GP baseline models, whilst offering the distinct advantage of inherent explainability. This transparency is crucial for fostering trust among clinical decision-makers, who often view opaque models with scepticism. Crucially, these models have demonstrated the capability to perform brain tumour segmentation in a weakly-supervised manner, eliminating the need for laborious pixel-wise annotations during training.

\subsection{Diagnostic accuracy versus interpretability}
\noindent In the domain of medical imaging, a marginal reduction in classification accuracy is often viewed as an acceptable trade-off for enhanced interpretability. The BraTS 2020 dataset experiments reveal that while the non-explainable baselines, InceptionV3 and ResNeXt50, achieved a marginally superior F1-score of 0.95, the GP-models remained highly competitive. Both GP-ShuffleUNet and GP-UNet achieved F1-scores of 0.93. A particularly revealing comparison arises with MProtoNet, a state-of-the-art interpretable baseline. Globally, MProtoNet demonstrated impressive performance, achieving an F1-score and accuracy of 0.97 (Table~\ref{Tab:MainResults3}). However, a more nuanced trend emerges when isolating the clinically critical tumour-only subset. As detailed in Table~\ref{Tab:MainResults5}, \emph{all} three proposed GP-models outperformed MProtoNet (97.79\%) in this targeted analysis. Specifically, GP-ShuffleUNet achieved a peak accuracy of 98.74\%, followed by GP-UNet (98.27\%) and GP-ReconResNet (97.99\%). This suggests that while MProtoNet is highly effective at distinguishing tumour-free slices, thereby boosting its global average, the GP-models possess superior discriminative power for classifying specific tumour subtypes once pathology is present. Thus, the GP-models offer a more robust solution for the precise grading of detected tumours.

\subsection{Segmentation efficacy and statistical significance}
A critical contribution of this work is the rigorous statistical validation of weakly-supervised segmentation performance. To benchmark the segmentation fidelity of the proposed GP-models, the paper compared them against MProtoNet \citep{wei2024mprotonet}, a state-of-the-art case-based interpretable classifier. It is important to note that MProtoNet is designed to provide transparency via prototypical part learning and was not originally engineered or evaluated for segmentation tasks. However, its architecture inherently generates attention maps to visualise reasoning. To facilitate a fair comparison in the weakly-supervised setting, binary segmentation masks were derived from these attribution maps using the same dynamic thresholding pipeline applied to the GP-models. 

The results indicate a statistically significant superiority of all GP-variants over the MProtoNet baseline ($p < 0.001$). Specifically, the GP-UNet achieved a median Dice score of 0.728 (95\% CI: 0.715--0.739). In contrast, the masks derived from MProtoNet yielded a median Dice score of 0.121. This substantial performance gap highlights a fundamental architectural distinction: whilst MProtoNet's prototype-based attention is sufficient for qualitative visual explanation, the global pooling mechanism integrated into the GP-models generates localisation heatmaps that correlate far more precisely with the morphological boundaries of the tumour. Furthermore, the raincloud plots (Fig.~\ref{fig:rainclouds}) illustrate a distinct distributional shift towards higher spatial accuracy for the GP-models, reinforcing their dual capability as both robust classifiers and effective weakly-supervised segmentors.

\subsection{Architectural nuances and dataset sensitivity}
A comparative analysis of the architectural variants across the two datasets reveals distinct behaviours regarding feature hierarchy and data efficiency. On the primary BraTS 2020 dataset, the U-Net-based variant (GP-UNet) emerged as the most effective architecture for weakly-supervised segmentation, achieving a median Dice score of 0.728. This can be attributed to two factors: firstly, the skip-connections inherent to the U-Net architecture facilitate the preservation of fine-grained spatial information crucial for localisation; secondly, the GP-UNet possesses the fewest trainable parameters (1.90 M), which likely acts as an implicit regulariser, mitigating overfitting on the complex, multi-modal MRI data.

Conversely, on the smaller Dataset \#1, the GP-ReconResNet demonstrated superior robustness, outperforming the GP-UNet in classification F1-score (0.95 vs 0.85). This inversion in performance suggests that the residual learning framework of the ResNet backbone may be more data-efficient and robust to geometric variations (such as mixed orientations) when training samples are scarce. However, this advantage diminishes when scaling to the larger, arguably more standardised BraTS dataset, where the lighter U-Net architecture prevails.

Regarding the stratification by tumour type, all models demonstrated superior segmentation performance on HGG compared to LGG. It is imperative to interpret this finding with caution. As HGG constitutes the vast majority of the training samples, the networks are inevitably biased towards the features of this dominant class. The performance disparity is therefore less likely an indicator of architectural affinity for aggressive tumour morphology, and more a reflection of the severe class imbalance. This hypothesis is statistically corroborated by the Precision-Recall analysis presented in Figure \ref{fig:PR}. The superior segmentation fidelity observed for the HGG subtype correlates directly with the class's near-perfect Average Precision (AP $>0.98$) across all architectures. This indicates that the networks successfully learn distinct, high-confidence features for the dominant HGG class, resulting in sharper, more accurate localisation heatmaps. Conversely, the degradation in LGG segmentation performance mirrors the precision decay observed in the LGG classification curves (e.g. GP-ReconResNet AP=0.897). This establishes a critical finding: in weakly-supervised frameworks, segmentation quality is intrinsically bound to the classifier's precision on the minority class. Consequently, the challenge is not merely one of pixel-wise delineation, but of optimising the decision boundary itself to be robust against class imbalance.

\subsection{Computational efficiency and training dynamics}
It must be acknowledged that GP-models require increased training time compared to non-GP models, primarily due to the expanding nature of the decoding pathway (Table~\ref{Tab:MainResults_para}). GP-ShuffleUNet, possessing the highest parameter count (26.4 M), necessitated the longest training duration ($\sim13$ days). In contrast, GP-UNet offers a pragmatic balance, delivering top-tier performance with only 1.90 M parameters and a significantly reduced training timeline ($\sim4$ days). Since GP-UNet minimises trainable parameters-thereby reducing the risk of overfitting, and maintains superior overall performance on the BraTS dataset, it may be considered the most efficient model for general deployment.

\subsection{Interpretability and trust}
Finally, regarding post-hoc interpretability, the GP-models are inherently explainable, whereas non-GP baselines (except MProtoNet) require post-hoc attribution techniques. Hence, a gradient-based technique - GBP and a feature-based technique - occlusion, were applied to all the models. For the GP-models, the interpretability results (also known as attention maps) were compared against the heatmaps of the networks - as they can be considered as a kind of "ground-truth". The models generally had reasonable attention maps, but there were some discrepancies and conflicting results between methods. However, assuming that the post-hoc methods are reliable (most of them agree with the networks' heatmaps), GP-UNet, GP-ShuffleUNet, and ResNext50 share similar outcomes, demonstrating proper tumour focal regions in both classes, followed by GP-ReconResNet, which has superior outcomes to InceptionV3. Even though InceptionV3 has the highest classification results (excluding MProtoNet), the attention maps show that positive and negative attributions are scattered compared to other models and often miss the tumour core. This reinforces the danger of relying on opaque models that may base decisions on spurious correlations, a phenomenon similarly observed in a COVID-19 classification study \citep{chatterjee2020exploration}. In contrast, GP-models constrain the decision-making process to the relevant anatomical regions, thereby fostering genuine clinical trust.
\section{Conclusion and Future Work}
\label{sec:conclusions}
\noindent This study addresses the critical dichotomy in medical imaging between high-performance "black-box" diagnostics and the clinical imperative for transparency, by transforming standard convolutional networks for segmentation and reconstruction tasks into inherently explainable classifiers. By integrating a Global Pooling mechanism, this paper demonstrated that architectures such as UNet, ShuffleUNet and ReconResNet can generate precise localisation heatmaps that drive classification decisions whilst simultaneously facilitating robust weakly-supervised segmentation. Whilst maintaining competitive global performance against opaque baselines, the proposed GP-models significantly outperformed all the baselines, including the interpretable MProtoNet on tumour-specific type classification. Furthermore, the comparative analysis of Receiver Operating Characteristic (ROC) and Precision-Recall (PR) curves highlighted the limitations of relying solely on global accuracy metrics in imbalanced datasets. Whilst ROC analysis suggested near-perfect separability, the PR curves exposed the specific challenges in characterising the minority LGG class, thereby validating that the models' superior performance on the dominant HGG subtype is driven by robust, high-confidence feature extraction rather than mere statistical optimisation. Crucially, the proposed framework revealed that all three models offer distinct operational advantages tailored to specific clinical needs. Whilst the GP-ShuffleUNet delivered unparalleled diagnostic precision-achieving a peak accuracy of 98.74\% on tumour-only images, the GP-UNet emerged as a highly efficient solution, balancing low computational cost with superior stability against class imbalance. Furthermore, the GP-ReconResNet demonstrated exceptional utility in handling complex data distributions; it not only exhibited superior data efficiency and robustness to geometric variations on the smaller, mixed-orientation dataset, but also significantly outperformed the U-Net variants in delineating the aggressive High-Grade Glioma (HGG) subtype. Collectively, these results confirm that the global pooling framework is architecture-agnostic, capable of leveraging the unique inductive biases of different backbones to outperform interpretable baselines like MProtoNet. 

Notwithstanding these advancements, the current framework operates on 2D slices, potentially discarding volumetric context, and entails higher computational demands than standard classifiers. Moreover, as highlighted by the Precision-Recall analysis, segmentation fidelity remains sensitive to severe class imbalance, necessitating future research into fully volumetric 3D architectures, model optimisation techniques (e.g. quantisation, pruning) to facilitate real-time inference, and advanced class-balancing strategies to decouple diagnostic precision from dataset distribution. 

Furthermore, the translation of these models into routine clinical workflows demands rigorous attention to the practical challenges of device heterogeneity and regulatory compliance. Future initiatives will therefore extend validation from retrospective in silico benchmarking to large-scale, multi-centre prospective trials and longitudinal studies, ensuring robustness against scanner variability whilst tracking disease progression over time. Simultaneously, adopting federated learning frameworks will be pivotal in training these data-hungry models across institutions without compromising patient privacy, thereby satisfying the stringent ethical and security prerequisites of Software as a Medical Device (SaMD) deployment.

Ultimately, this work offers a versatile foundation for trustworthy clinical decision support. Beyond these immediate architectural refinements, a particularly promising avenue lies in integrating the generated localisation priors with Vision-Language Models (VLMs) to facilitate automated, biologically-grounded radiological reporting. Finally, extending the Global Pooling mechanism to heterogeneous multi-modal sources, integrating genomics or clinical health records, will be essential in establishing this framework as a truly agnostic paradigm for explainable, holistic medical diagnostics.

\section*{Acknowledgement}
This work was in part conducted within the context of the International Graduate School MEMoRIAL at Otto von Guericke
University (OVGU) Magdeburg, Germany, kindly supported by the European Structural and Investment Funds (ESF) under the
programme "Sachsen-Anhalt WISSENSCHAFT Internationalisierung" (project no. ZS/2016/08/80646).

\bibliography{mybibfile}

@inproceedings{hasan2023explainable,
  title={Explainable Automated Brain Tumor Detection Using CNN},
  author={Hasan, Mohammad Asif and Sarker, Hasan and Goni, Md Omaer Faruq},
  booktitle={International Conference on Big Data, IoT and Machine Learning},
  pages={481--496},
  year={2023},
  organization={Springer}
}

@inproceedings{wei2024mprotonet,
  title={Mprotonet: A case-based interpretable model for brain tumor classification with 3d multi-parametric magnetic resonance imaging},
  author={Wei, Yuanyuan and Tam, Roger and Tang, Xiaoying},
  booktitle={Medical Imaging with Deep Learning},
  pages={1798--1812},
  year={2024},
  organization={PMLR}
}

@article{chen2023ws,
  title={WS-MTST: Weakly supervised multi-label brain tumor segmentation with transformers},
  author={Chen, Huazhen and An, Jianpeng and Jiang, Bochang and Xia, Lili and Bai, Yunhao and Gao, Zhongke},
  journal={IEEE Journal of Biomedical and Health Informatics},
  volume={27},
  number={12},
  pages={5914--5925},
  year={2023},
  publisher={IEEE}
}

@ARTICLE{han2025channel,
  author={Han, Yan and Liu, Kai and Yuan, Lingling and Rahaman, Md and Grzegorzek, Marcin and Sun, Hongzan and Li, Chen and Chen, Huiling},
  journal={IEEE Journal of Biomedical and Health Informatics}, 
  title={Channel-Gated Transformers With Affinity CAM for Weakly Supervised Multi-Class Brain Tumor Segmentation}, 
  year={2025},
  volume={},
  number={},
  pages={1-14},
  doi={10.1109/JBHI.2025.3634736}}

@article{chen2025dual,
  title={Dual degradation image inpainting method via adaptive feature fusion and U-net network},
  author={Chen, Yuantao and Xia, Runlong and Yang, Kai and Zou, Ke},
  journal={Applied Soft Computing},
  volume={174},
  pages={113010},
  year={2025},
  publisher={Elsevier}
}

@article{zhang2025atm,
  title={ATM-DEN: Image Inpainting via attention transfer module and Decoder-Encoder network},
  author={Zhang, Siwei and Chen, Yuantao},
  journal={Signal Processing: Image Communication},
  volume={133},
  pages={117268},
  year={2025},
  publisher={Elsevier}
}

@article{zhang2025mgnet,
  title={MGNet: RGBT tracking via cross-modality cross-region mutual guidance},
  author={Zhang, Jianming and Yang, Jing and Qin, Yu and Xiao, Zhu and Wang, Jin},
  journal={Neural Networks},
  pages={107707},
  year={2025},
  publisher={Elsevier}
}

@article{zhang2025rgbt,
  title={RGBT tracking via frequency-aware feature enhancement and unidirectional mixed attention},
  author={Zhang, Jianming and Yang, Jing and Liu, Zikang and Wang, Jin},
  journal={Neurocomputing},
  volume={616},
  pages={128908},
  year={2025},
  publisher={Elsevier}
}

@article{singh2025cicada,
  title={CICADA (UCX): A novel approach for automated breast cancer classification through aggressiveness delineation},
  author={Singh, Davinder Paul and Banerjee, Tathagat and Kour, Pawandeep and Swain, Debabrata and Narayan, Yogendra},
  journal={Computational Biology and Chemistry},
  volume={115},
  pages={108368},
  year={2025},
  publisher={Elsevier}
}

@article{singh2025comprehensive,
  title={A comprehensive study of enhanced computational approaches for breast cancer classification: comparative analysis with existing state of the art methods},
  author={Singh, Davinder Paul and Banerjee, Tathagat and Kour, Pawandeep and Malik, Rahul and Naidu, Gangu Rama and Kumar, Raju and Kaushal, Rajanish Kumar and Singh, Ram Murat and Narayan, Yogendra and others},
  journal={Archives of Computational Methods in Engineering},
  pages={1--29},
  year={2025},
  publisher={Springer}
}

@article{singh2025comprehensiveratino,
  title={A comprehensive study on deep learning models for the detection of diabetic retinopathy using pathological images},
  author={Singh, Davinder Paul and Banerjee, Tathagat and Mahajan, Shubham and Ramesh Chandra, K and Kumar, Raju and Phani, Shanta and Kour, Pawandeep and Narayan, Yogendra and Kour, Gurpreet and Mukhija, Loveleena and others},
  journal={Archives of Computational Methods in Engineering},
  pages={1--30},
  year={2025},
  publisher={Springer}
}

@article{banerjee2025advances,
  title={Advances in deep Neural, transformer Learning, and Kernel-Based methods for diabetic retinopathy detection: A comprehensive review},
  author={Banerjee, Tathagat and Singh, Davinder Paul and Kour, Pawandeep},
  journal={Archives of Computational Methods in Engineering},
  pages={1--49},
  year={2025},
  publisher={Springer}
}

@article{banerjee2025towards,
  title={Towards automated and reliable lung cancer detection in histopathological images using DY-FSPAN: A feature-summarized pyramidal attention network for explainable AI},
  author={Banerjee, Tathagat},
  journal={Computational Biology and Chemistry},
  pages={108500},
  year={2025},
  publisher={Elsevier}
}

@article{banerjee2025comparing,
  title={Comparing bipartite convoluted and attention-driven methods for skin cancer detection: A review of explainable AI and transfer learning strategies},
  author={Banerjee, Tathagat},
  journal={Archives of Computational Methods in Engineering},
  pages={1--25},
  year={2025},
  publisher={Springer}
}

@article{banerjee2025electromagnetic,
  title={Electromagnetic Interaction Algorithm (EIA)-Based Feature Selection With Adaptive Kernel Attention Network (AKAttNet) for Autism Spectrum Disorder Classification},
  author={Banerjee, Tathagat},
  journal={International Journal of Developmental Neuroscience},
  volume={85},
  number={5},
  pages={e70034},
  year={2025},
  publisher={Wiley Online Library}
}

@article{banerjee2025pyramidal,
  title={Pyramidal attention-based T network for brain tumor classification: a comprehensive analysis of transfer learning approaches for clinically reliable and reliable AI hybrid approaches},
  author={Banerjee, Tathagat and Chhabra, Prachi and Kumar, Manoj and Kumar, Abhay and Abhishek, Kumar and Shah, Mohd Asif},
  journal={Scientific Reports},
  volume={15},
  number={1},
  pages={28669},
  year={2025},
  publisher={Nature Publishing Group UK London}
}

@article{banerjee2025novel,
  title={A novel hybrid deep learning approach combining deep feature attention and statistical validation for enhanced thyroid ultrasound segmentation},
  author={Banerjee, Tathagat and Singh, Davinder Paul and Swain, Debabrata and Mahajan, Shubham and Kadry, Seifedine and Kim, Jungeun},
  journal={Scientific Reports},
  volume={15},
  number={1},
  pages={27207},
  year={2025},
  publisher={Nature Publishing Group UK London}
}

@article{narayan2025comparative,
  title={A comparative evaluation of deep learning architectures for prostate cancer segmentation: Introducing TrionixNet with N-core multi-attention mechanism},
  author={Narayan, Yogendra and Singh, Davinder Paul and Banerjee, Tathagat and Kour, Pawandeep and Rane, Kantilal and C, Anand Deva Durai and Chandar, K Punnam and Kaushal, Rajanish Kumar and Singh, Ram Murat and Dhillon, Yuvika},
  journal={Archives of Computational Methods in Engineering},
  pages={1--40},
  year={2025},
  publisher={Springer}
}

@article{pacal2026towards,
  title={Towards accurate and interpretable brain tumor diagnosis: T-FSPANNet with tri-attribute and pyramidal attention-based feature fusion},
  author={Pacal, Ishak and Banerjee, Tathagat},
  journal={Biomedical Signal Processing and Control},
  volume={113},
  pages={108852},
  year={2026},
  publisher={Elsevier}
}

@article{banerjee2025UIGO,
  title={A novel unified Inception-U-Net hybrid gravitational optimization model (UIGO) incorporating automated medical image segmentation and feature selection for liver tumor detection},
  author={Banerjee, Tathagat and Singh, Davinder Paul and Kour, Pawandeep and Swain, Debabrata and Mahajan, Shubham and Kadry, Seifedine and Kim, Jungeun},
  journal={Scientific Reports},
  volume={15},
  number={1},
  pages={29908},
  year={2025},
  publisher={Nature Publishing Group UK London}
}

@article{hafeez2023cnn,
  title={A CNN-Model to Classify Low-grade and High-grade Glioma from MRI Images},
  author={Hafeez, Hafiz Aamir and Elmagzoub, MOHAMED A and Abdullah, Nurul Azma Binti and Al Reshan, Mana Saleh and Gilanie, Ghulam and Alyami, Sultan and ul Hassan, Mahmood and Shaikh, Asadullah},
  journal={IEEE Access},
  year={2023},
  publisher={IEEE}
}

@article{dutta2024arm,
  title={ARM-Net: Attention-guided residual multiscale CNN for multiclass brain tumor classification using MR images},
  author={Dutta, Tapas Kumar and Nayak, Deepak Ranjan and Zhang, Yu-Dong},
  journal={Biomedical Signal Processing and Control},
  volume={87},
  pages={105421},
  year={2024},
  publisher={Elsevier}
}

@article{barstugan2023classification,
  title={Classification of 3D-DWT Features of Brain Tumours with SVM},
  author={Barstugan, Mucahid},
  journal={Orclever Proceedings of Research and Development},
  volume={2},
  number={1},
  pages={39--49},
  year={2023}
}

@article{Badza.2020,
 author = {Bad{\v{z}}a, Milica M. and Barjaktarovi{\'c}, Marko {\v{C}}.},
 year = {2020},
 title = {Classification of Brain Tumors from MRI Images Using a Convolutional Neural Network},
 pages = {1999},
 volume = {10},
 number = {6},
 journal = {Applied Sciences},
 doi = {10.3390/app10061999}
}

@inproceedings{chatterjee2021shuffleunet,
  title={ShuffleUNet: Super resolution of diffusion-weighted MRIs using deep learning},
  author={Chatterjee, Soumick and Sciarra, Alessandro and D{\"u}nnwald, Max and Mushunuri, Raghava Vinaykanth and Podishetti, Ranadheer and Rao, Rajatha Nagaraja and Gopinath, Geetha Doddapaneni and Oeltze-Jafra, Steffen and Speck, Oliver and N{\"u}rnberger, Andreas},
  booktitle={2021 29th European Signal Processing Conference (EUSIPCO)},
  pages={940--944},
  year={2021},
  organization={IEEE},
  doi={10.23919/EUSIPCO54536.2021.9615963}
}

@inproceedings{Dubost.5222017,
  title={Gp-unet: Lesion detection from weak labels with a 3d regression network},
  author={Dubost, Florian and Bortsova, Gerda and Adams, Hieab and Ikram, Arfan and Niessen, Wiro J and Vernooij, Meike and Bruijne, Marleen De},
  booktitle={International Conference on Medical Image Computing and Computer-Assisted Intervention},
  pages={214--221},
  year={2017},
  organization={Springer}
}

@article{Irmak.2021,
 author = {Irmak, Emrah},
 year = {2021},
 title = {Multi-Classification of Brain Tumor MRI Images Using Deep Convolutional Neural Network with Fully Optimized Framework},
 pages = {1015--1036},
 volume = {45},
 number = {3},
 issn = {2228-6179},
 journal = {Iranian Journal of Science and Technology, Transactions of Electrical Engineering},
 doi = {10.1007/s40998-021-00426-9}
}

@incollection{Noor.2019,
 author = {Noor, Manan Binth Taj and Zenia, Nusrat Zerin and Kaiser, M. Shamim and Mahmud, Mufti and {Al Mamun}, Shamim},
 title = {Detecting Neurodegenerative Disease from MRI: A Brief Review on a Deep Learning Perspective},
 pages = {115--125},
 volume = {11976},
 publisher = {{Springer International Publishing}},
 isbn = {978-3-030-37077-0},
 series = {Lecture Notes in Computer Science},
 editor = {Liang, Peipeng and Goel, Vinod and Shan, Chunlei},
 booktitle = {Brain Informatics},
 year = {2019},
 address = {Cham},
 doi = {10.1007/978-3-030-37078-7{\textunderscore }12}
}

@inproceedings{Ronneberger.5182015,
  title={U-net: Convolutional networks for biomedical image segmentation},
  author={Ronneberger, Olaf and Fischer, Philipp and Brox, Thomas},
  booktitle={International Conference on Medical image computing and computer-assisted intervention},
  pages={234--241},
  year={2015},
  organization={Springer}
}

@article{menze2014multimodal,
  title={The multimodal brain tumor image segmentation benchmark (BRATS)},
  author={Menze, Bjoern H and Jakab, Andras and Bauer, Stefan and Kalpathy-Cramer, Jayashree and Farahani, Keyvan and Kirby, Justin and Burren, Yuliya and Porz, Nicole and Slotboom, Johannes and Wiest, Roland and others},
  journal={IEEE transactions on medical imaging},
  volume={34},
  number={10},
  pages={1993--2024},
  year={2014},
  publisher={IEEE}
}

@article{lloyd2017high,
  title={High resolution global gridded data for use in population studies},
  author={Lloyd, Christopher T and Sorichetta, Alessandro and Tatem, Andrew J},
  journal={Scientific data},
  volume={4},
  number={1},
  pages={1--17},
  year={2017},
  publisher={Nature Publishing Group}
}

@inproceedings{latif2017multiclass,
  title={Multiclass brain Glioma tumor classification using block-based 3D Wavelet features of MR images},
  author={Latif, Ghazanfar and Butt, M Mohsin and Khan, Adil H and Butt, Omair and Iskandar, DNF Awang},
  booktitle={2017 4th International Conference on Electrical and Electronic Engineering (ICEEE)},
  pages={333--337},
  year={2017},
  organization={IEEE}
}

@inproceedings{cho2017classification,
  title={Classification of low-grade and high-grade glioma using multi-modal image radiomics features},
  author={Cho, Hwan-ho and Park, Hyunjin},
  booktitle={2017 39th Annual International Conference of the IEEE Engineering in Medicine and Biology Society (EMBC)},
  pages={3081--3084},
  year={2017},
  organization={IEEE}
}

@inproceedings{szegedy2016rethinking,
  title={Rethinking the inception architecture for computer vision},
  author={Szegedy, Christian and Vanhoucke, Vincent and Ioffe, Sergey and Shlens, Jon and Wojna, Zbigniew},
  booktitle={Proceedings of the IEEE conference on computer vision and pattern recognition},
  pages={2818--2826},
  year={2016}
}

@inproceedings{xie2017aggregated,
  title={Aggregated residual transformations for deep neural networks},
  author={Xie, Saining and Girshick, Ross and Doll{\'a}r, Piotr and Tu, Zhuowen and He, Kaiming},
  booktitle={Proceedings of the IEEE conference on computer vision and pattern recognition},
  pages={1492--1500},
  year={2017}
}

@article{bakas2018identifying,
  title={Identifying the best machine learning algorithms for brain tumor segmentation, progression assessment, and overall survival prediction in the BRATS challenge},
  author={Bakas, Spyridon and Reyes, Mauricio and Jakab, Andras and Bauer, Stefan and Rempfler, Markus and Crimi, Alessandro and Shinohara, Russell Takeshi and Berger, Christoph and Ha, Sung Min and Rozycki, Martin and others},
  journal={arXiv preprint arXiv:1811.02629},
  year={2018}
}

@article{Sultan.2019,
 author = {Sultan, Hossam H. and Salem, Nancy M. and Al-Atabany, Walid},
 year = {2019},
 title = {Multi-Classification of Brain Tumor Images Using Deep Neural Network},
 pages = {69215--69225},
 volume = {7},
 journal = {IEEE Access},
 doi = {10.1109/ACCESS.2019.2919122}
}

@article{chatterjee2020exploration,
  title={Exploration of interpretability techniques for deep covid-19 classification using chest x-ray images},
  author={Chatterjee, Soumick and Saad, Fatima and Sarasaen, Chompunuch and Ghosh, Suhita and Khatun, Rupali and Radeva, Petia and Rose, Georg and Stober, Sebastian and Speck, Oliver and N{\"u}rnberger, Andreas},
  journal={arXiv preprint arXiv:2006.02570},
  year={2020}
}

@misc{JunCheng.2017,
 author = {{Jun Cheng}},
 date = {2017},
 title = {brain tumor dataset},
 publisher = {figshare},
 doi = {10.6084/m9.figshare.1512427.v5}
}

@article{krizhevsky2012imagenet,
  title={Imagenet classification with deep convolutional neural networks},
  author={Krizhevsky, Alex and Sutskever, Ilya and Hinton, Geoffrey E},
  journal={Advances in neural information processing systems},
  volume={25},
  pages={1097--1105},
  year={2012}
}

@misc{Wong.9282016,
 author = {Wong, Sebastien C. and Gatt, Adam and Stamatescu, Victor and McDonnell, Mark D.},
 date = {9/28/2016},
 title = {Understanding data augmentation for classification: when to warp?},
 url = {http://arxiv.org/pdf/1609.08764v2}
 }

@article{Jeong.2014,
 author = {Jeong, Boseul and Choi, Dae Seob and Shin, Hwa Seon and Choi, Hye Young and Park, Mi Jung and Jeon, Kyung Nyeo and Na, Jae Beom and Chung, Sung Hoon},
 year = {2014},
 title = {T1-weighted FLAIR MR Imaging for the Evaluation of Enhancing Brain Tumors: Comparison with Spin Echo Imaging},
 pages = {151},
 volume = {18},
 number = {2},
 issn = {1226-9751},
 journal = {Journal of the Korean Society of Magnetic Resonance in Medicine},
 doi = {10.13104/jksmrm.2014.18.2.151}
}

@article{Ayadi.2021,
 author = {Ayadi, Wadhah and Elhamzi, Wajdi and Charfi, Imen and Atri, Mohamed},
 year = {2021},
 title = {Deep CNN for Brain Tumor Classification},
 pages = {671--700},
 volume = {53},
 number = {1},
 issn = {1370-4621},
 journal = {Neural Processing Letters},
 doi = {10.1007/s11063-020-10398-2}
}

@incollection{Abiwinanda.2019,
 author = {Abiwinanda, Nyoman and Hanif, Muhammad and Hesaputra, S. Tafwida and Handayani, Astri and Mengko, Tati Rajab},
 title = {Brain Tumor Classification Using Convolutional Neural Network},
 pages = {183--189},
 volume = {68/1},
 publisher = {{Springer Singapore}},
 isbn = {978-981-10-9034-9},
 series = {IFMBE Proceedings},
 editor = {Lhotska, Lenka and Sukupova, Lucie and Lackovi{\'c}, Igor and Ibbott, Geoffrey S.},
 booktitle = {World Congress on Medical Physics and Biomedical Engineering 2018},
 year = {2019},
 address = {Singapore},
 doi = {10.1007/978-981-10-9035-6{\textunderscore }33}
}

@article{DiazPernas.2021,
 author = {D{\'i}az-Pernas, Francisco Javier and Mart{\'i}nez-Zarzuela, Mario and Ant{\'o}n-Rodr{\'i}guez, M{\'i}riam and Gonz{\'a}lez-Ortega, David},
 year = {2021},
 title = {A Deep Learning Approach for Brain Tumor Classification and Segmentation Using a Multiscale Convolutional Neural Network},
 volume = {9},
 number = {2},
 issn = {2227-9032},
 journal = {Healthcare (Basel, Switzerland)},
 doi = {10.3390/healthcare9020153}
}

@article{Li.2019,
 author = {Li, Haichun and Li, Ao and Wang, Minghui},
 year = {2019},
 title = {A novel end-to-end brain tumor segmentation method using improved fully convolutional networks},
 pages = {150--160},
 volume = {108},
 journal = {Computers in biology and medicine}
}

@article{Pereira.2016,
 author = {Pereira, S{\'e}rgio and Pinto, Adriano and Alves, Victor and Silva, Carlos A.},
 year = {2016},
 title = {Brain tumor segmentation using convolutional neural networks in MRI images},
 pages = {1240--1251},
 volume = {35},
 number = {5},
 journal = {IEEE transactions on medical imaging}
}

@article{Razzak.2018,
 author = {Razzak, Muhammad Imran and Imran, Muhammad and Xu, Guandong},
 year = {2018},
 title = {Efficient brain tumor segmentation with multiscale two-pathway-group conventional neural networks},
 pages = {1911--1919},
 volume = {23},
 number = {5},
 journal = {IEEE journal of biomedical and health informatics}
}

@article{Deepak.2019,
  title={Brain tumor classification using deep CNN features via transfer learning},
  author={Deepak, S and Ameer, PM},
  journal={Computers in biology and medicine},
  volume={111},
  pages={103345},
  year={2019},
  publisher={Elsevier}
}

@article{Khawaldeh.2018,
 author = {Khawaldeh, Saed and Pervaiz, Usama and Rafiq, Azhar and Alkhawaldeh, Rami S.},
 year = {2018},
 title = {Noninvasive grading of glioma tumor using magnetic resonance imaging with convolutional neural networks},
 pages = {27},
 volume = {8},
 number = {1},
 journal = {Applied Sciences}
}

@article{Rehman.2020,
 author = {Rehman, Arshia and Naz, Saeeda and Razzak, Muhammad Imran and Akram, Faiza and Imran, Muhammad},
 year = {2020},
 title = {A deep learning-based framework for automatic brain tumors classification using transfer learning},
 pages = {757--775},
 volume = {39},
 number = {2},
 journal = {Circuits, Systems, and Signal Processing}
}

@article{chatterjee2021spatiotemp,
  title={Classification of brain tumours in MR images using deep spatiospatial models},
  author={Chatterjee, Soumick and Nizamani, Faraz Ahmed and N{\"u}rnberger, Andreas and Speck, Oliver},
  journal={Scientific Reports},
  volume={12},
  number={1},
  pages={1--11},
  year={2022},
  publisher={Nature Publishing Group},
  doi={10.1038/s41598-022-05572-6}
}

@article{chatterjee2021torchesegeta,
  title={TorchEsegeta: Framework for Interpretability and Explainability of Image-based Deep Learning Models},
  author={Chatterjee, Soumick and Das, Arnab and Mandal, Chirag and Mukhopadhyay, Budhaditya and Vipinraj, Manish and Shukla, Aniruddh and Nagaraja Rao, Rajatha and Sarasaen, Chompunuch and Speck, Oliver and N{\"u}rnberger, Andreas},
  journal={Applied Sciences},
  volume={12},
  number={4},
  pages={1834},
  year={2022},
  publisher={Multidisciplinary Digital Publishing Institute},
  doi={10.3390/app12041834}
}

@article{chatterjee2021reconresnet,
  title={ReconResNet: Regularised residual learning for MR image reconstruction of Undersampled Cartesian and Radial data},
  author={Chatterjee, Soumick and Breitkopf, Mario and Sarasaen, Chompunuch and Yassin, Hadya and Rose, Georg and N{\"u}rnberger, Andreas and Speck, Oliver},
  journal={Computers in Biology and Medicine},
  pages={105321},
  year={2022},
  publisher={Elsevier},
  doi={10.1016/j.compbiomed.2022.105321}
}

@inproceedings{shahzadi2018cnn,
  title={CNN-LSTM: Cascaded framework for brain tumour classification},
  author={Shahzadi, Iram and Tang, Tong Boon and Meriadeau, Fabrice and Quyyum, Abdul},
  booktitle={2018 IEEE-EMBS Conference on Biomedical Engineering and Sciences (IECBES)},
  pages={633--637},
  year={2018},
  organization={IEEE}
}

@inproceedings{ge2018deep,
  title={Deep learning and multi-sensor fusion for glioma classification using multistream 2D convolutional networks},
  author={Ge, Chenjie and Gu, Irene Yu-Hua and Jakola, Asgeir Store and Yang, Jie},
  booktitle={2018 40th Annual International Conference of the IEEE Engineering in Medicine and Biology Society (EMBC)},
  pages={5894--5897},
  year={2018},
  organization={IEEE}
}

@article{yang2018glioma,
  title={Glioma grading on conventional MR images: a deep learning study with transfer learning},
  author={Yang, Yang and Yan, Lin-Feng and Zhang, Xin and Han, Yu and Nan, Hai-Yan and Hu, Yu-Chuan and Hu, Bo and Yan, Song-Lin and Zhang, Jin and Cheng, Dong-Liang and others},
  journal={Frontiers in neuroscience},
  pages={804},
  year={2018},
  publisher={Frontiers}
}

@article{zhuge2020automated,
  title={Automated glioma grading on conventional MRI images using deep convolutional neural networks},
  author={Zhuge, Ying and Ning, Holly and Mathen, Peter and Cheng, Jason Y and Krauze, Andra V and Camphausen, Kevin and Miller, Robert W},
  journal={Medical physics},
  volume={47},
  number={7},
  pages={3044--3053},
  year={2020},
  publisher={Wiley Online Library}
}

@article{mzoughi2020deep,
  title={Deep multi-scale 3D convolutional neural network (CNN) for MRI gliomas brain tumor classification},
  author={Mzoughi, Hiba and Njeh, Ines and Wali, Ali and Slima, Mohamed Ben and BenHamida, Ahmed and Mhiri, Chokri and Mahfoudhe, Kharedine Ben},
  journal={Journal of Digital Imaging},
  volume={33},
  number={4},
  pages={903--915},
  year={2020},
  publisher={Springer}
}

@article{KENNY2021103459,
title = {Explaining black-box classifiers using post-hoc explanations-by-example: The effect of explanations and error-rates in XAI user studies},
journal = {Artificial Intelligence},
volume = {294},
pages = {103459},
year = {2021},
issn = {0004-3702},
doi = {https://doi.org/10.1016/j.artint.2021.103459},
url = {https://www.sciencedirect.com/science/article/pii/S0004370221000102},
author = {Eoin M. Kenny and Courtney Ford and Molly Quinn and Mark T. Keane}
}

@article{DBLP:journals/corr/abs-2108-04840,
  author    = {Andreas Madsen and
               Siva Reddy and
               Sarath Chandar},
  title     = {Post-hoc Interpretability for Neural {NLP:} {A} Survey},
  journal   = {CoRR},
  volume    = {abs/2108.04840},
  year      = {2021},
  url       = {https://arxiv.org/abs/2108.04840},
  eprinttype = {arXiv},
  eprint    = {2108.04840},
  bibsource = {dblp computer science bibliography, https://dblp.org}
}

@phdthesis{laugel2020local,
  title={Local Post-hoc Interpretability for Black-box Classifiers},
  author={Laugel, Thibault},
  year={2020},
  school={Sorbonne Universit{\'e}, CNRS, LIP6, F-75005 Paris, France}
}

\appendix
\clearpage
\section*{Appendix}
\begin{figure}[htbp]
    \centering
    \includegraphics[width=0.48\textwidth]{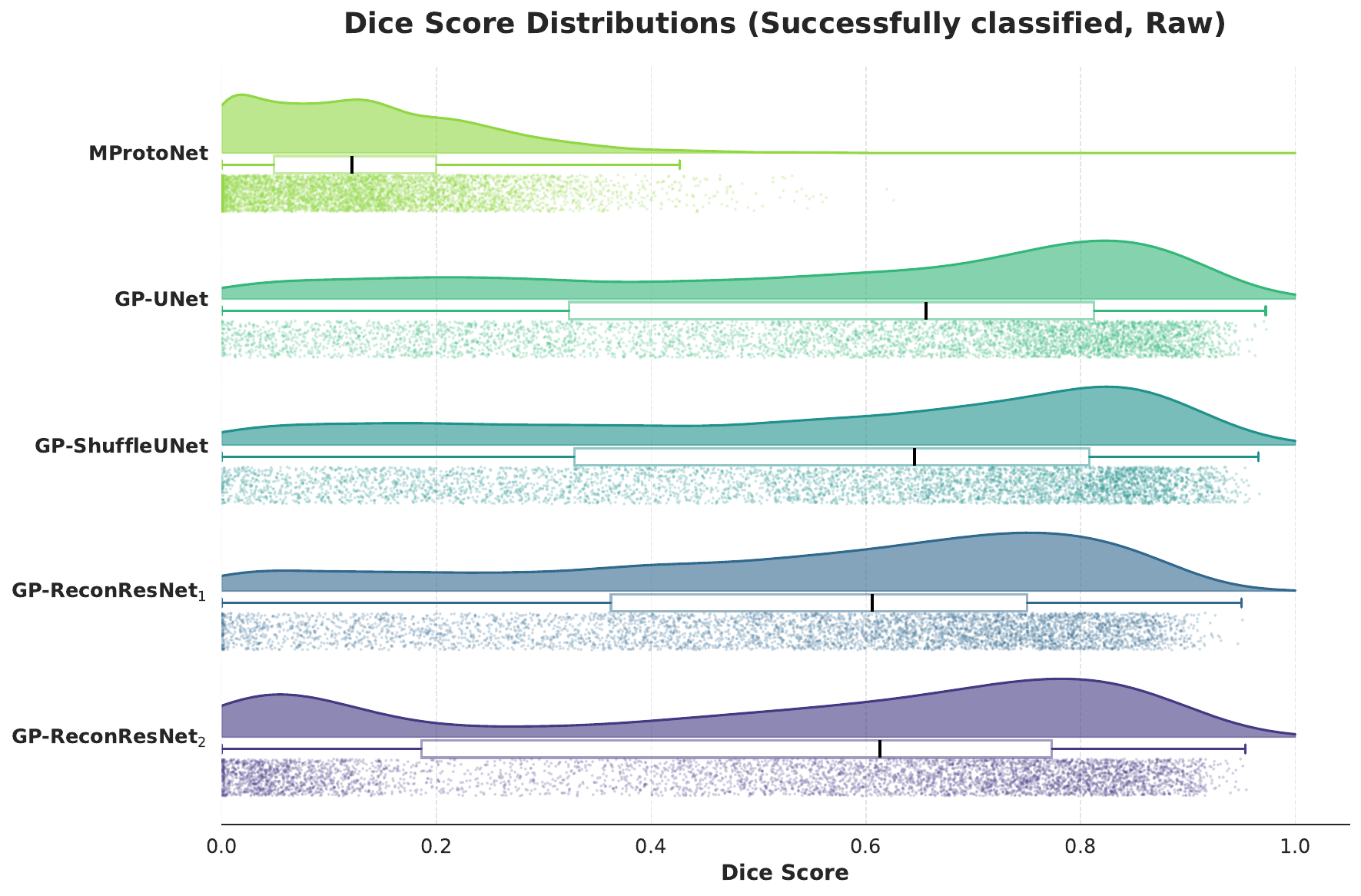}
    \caption{Distribution of Dice scores for the subset of successfully classified images. In contrast to Figure~\ref{fig:rainclouds}a, this plot illustrates the segmentation fidelity prior to the application of post-processing. Comparison with the main text reveals the specific quantitative gain attributed to the post-processing stage.}
    \label{fig:app_success_raw}
\end{figure}

\begin{figure}[htbp]
    \centering
    \includegraphics[width=0.48\textwidth]{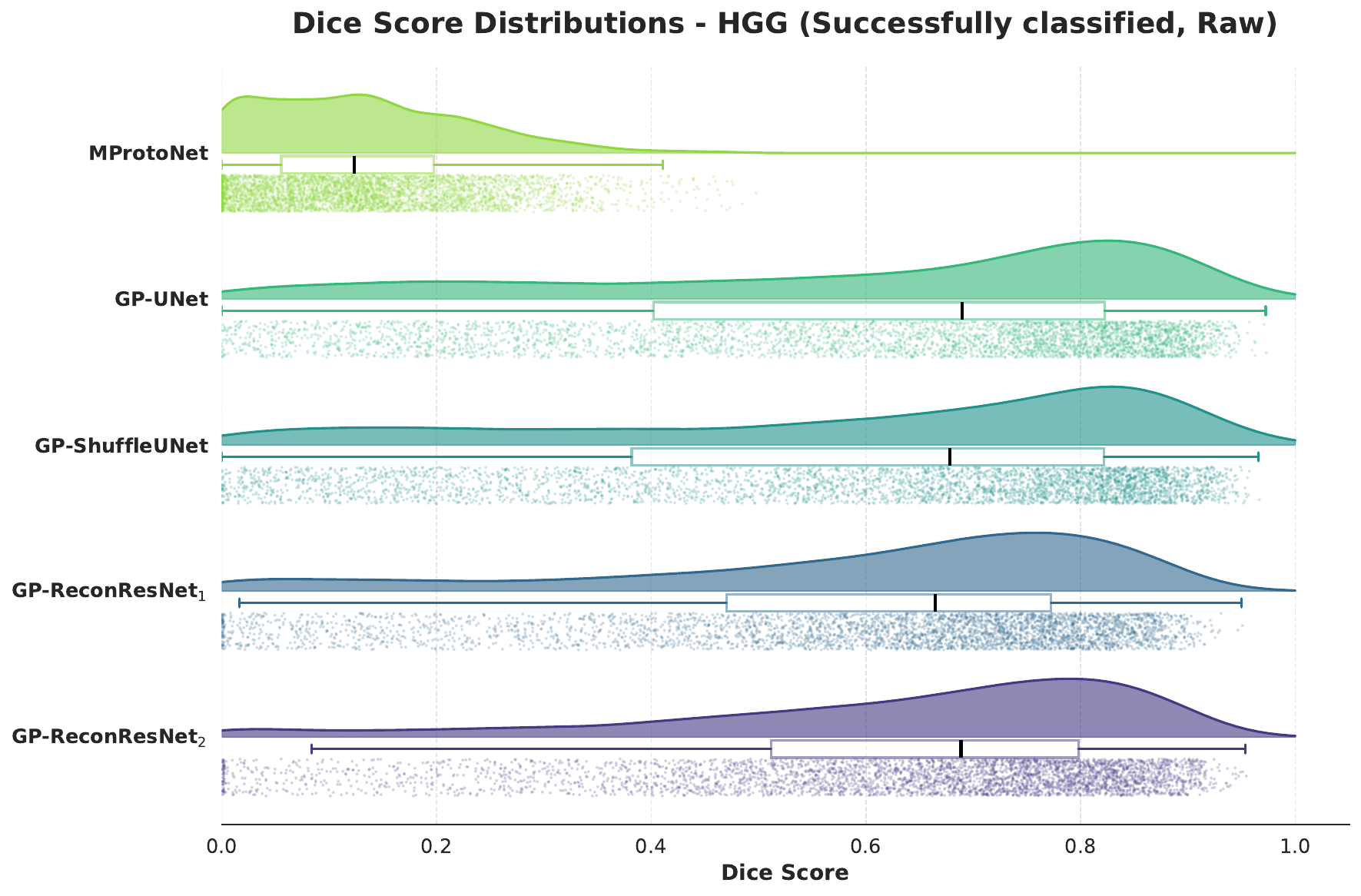}
    \caption{Distribution of Dice scores for the successfully classified HGG subset. This serves as the pre-processing baseline for the HGG analysis presented in Figure~\ref{fig:rainclouds}b, highlighting the intrinsic performance of the architectures on this predominant pathological class.}
    \label{fig:app_hgg_success_raw}
\end{figure}

\begin{figure}[htbp]
    \centering
    \includegraphics[width=0.48\textwidth]{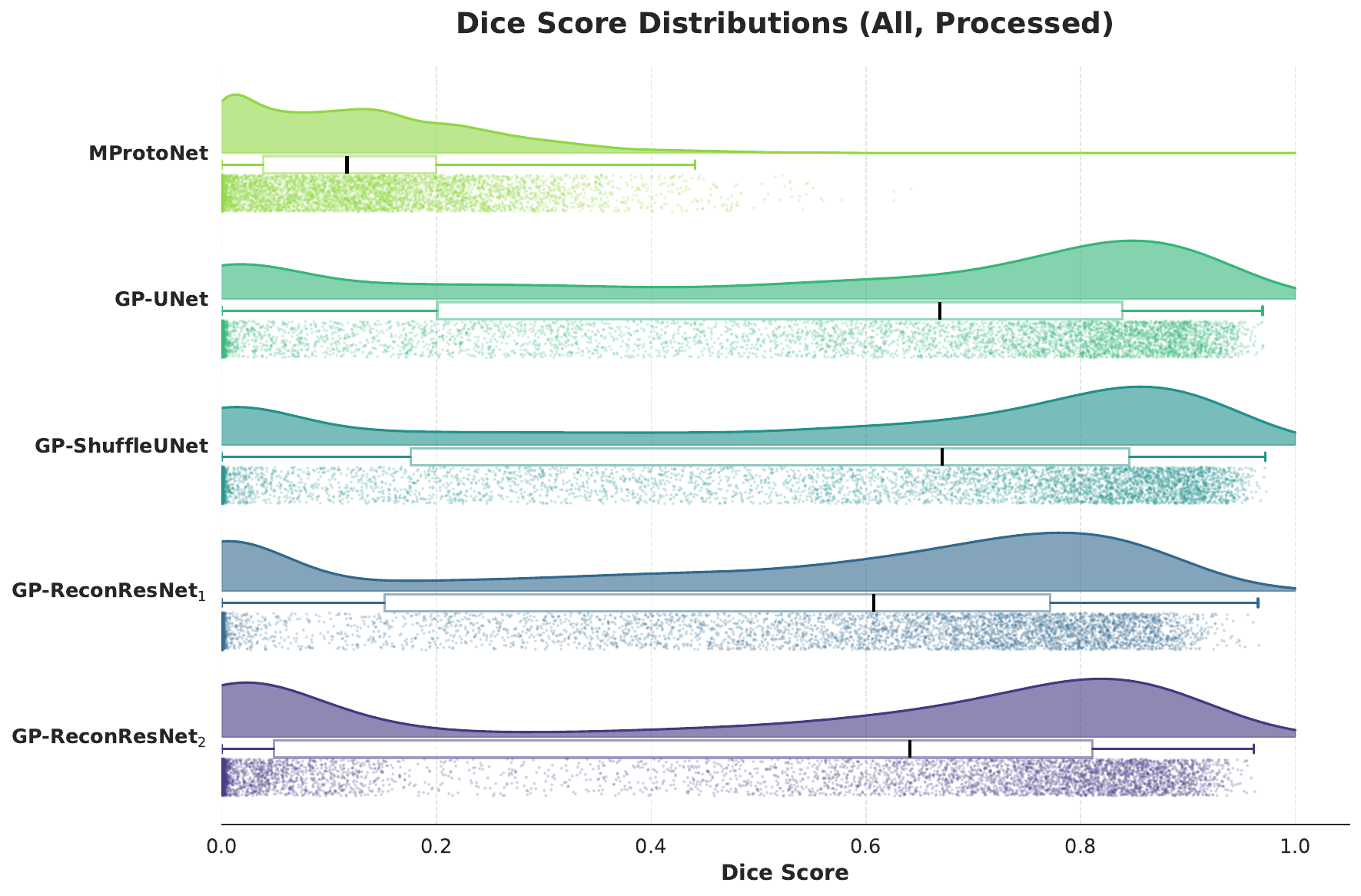}
    \caption{Distribution of post-processed Dice scores across the complete dataset. This visualises model performance on the full cohort, inclusive of cases that were excluded from the "successfully classified" subset. It assesses the generalisation capability of the proposed architectures when no filtration for classification success is applied.}
    \label{fig:app_all_processed}
\end{figure}

\FloatBarrier
\begin{figure}[htbp]
    \centering
    \includegraphics[width=0.48\textwidth]{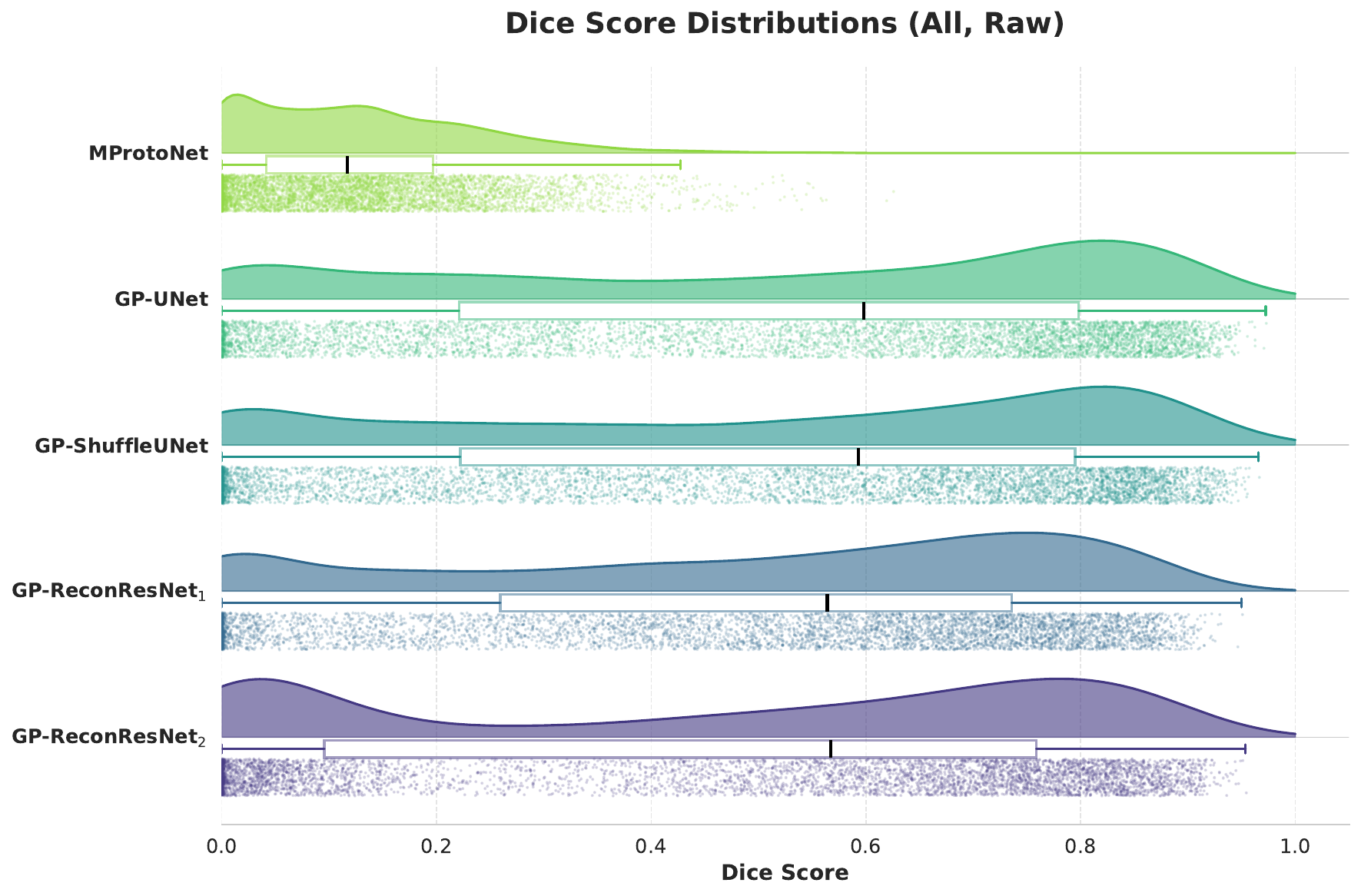}
    \caption{Distribution of raw Dice scores across the complete dataset. This plot depicts the foundational segmentation fidelity of the GP-models prior to any post-processing steps or exclusion criteria, providing a global baseline for the study.}
    \label{fig:app_all_raw}
\end{figure}

\begin{figure}[htbp]
    \centering
    \includegraphics[width=0.48\textwidth]{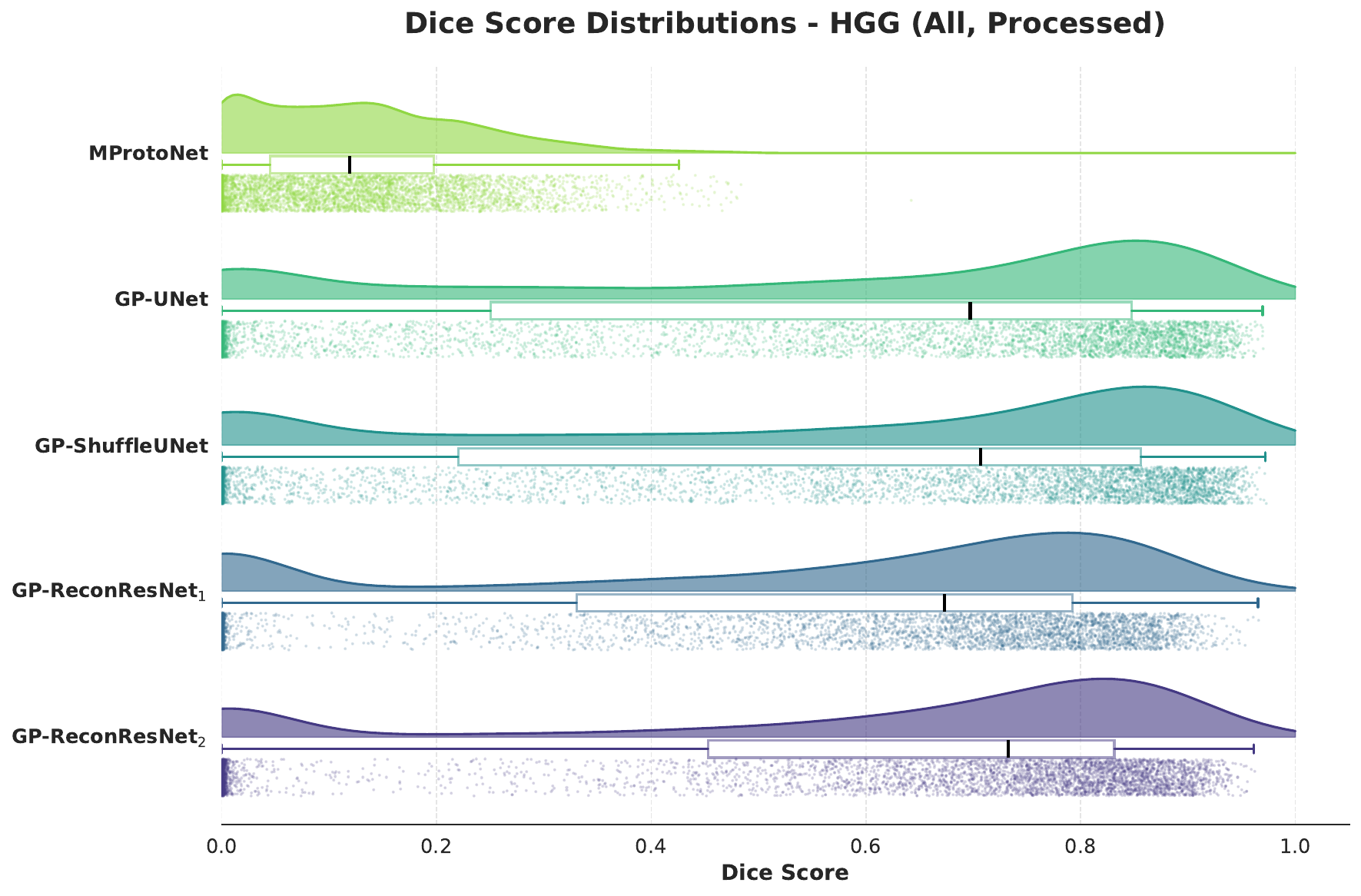}
    \caption{Stratified analysis of post-processed Dice scores for the HGG subtype across the full dataset. This demonstrates the robustness of the reconstruction-based methods (specifically GP-ReconResNet$_2$) on HGG cases even when classification failures are included in the evaluation.}
    \label{fig:app_hgg_all_processed}
\end{figure}

\begin{figure}[htbp]
    \centering
    \includegraphics[width=0.48\textwidth]{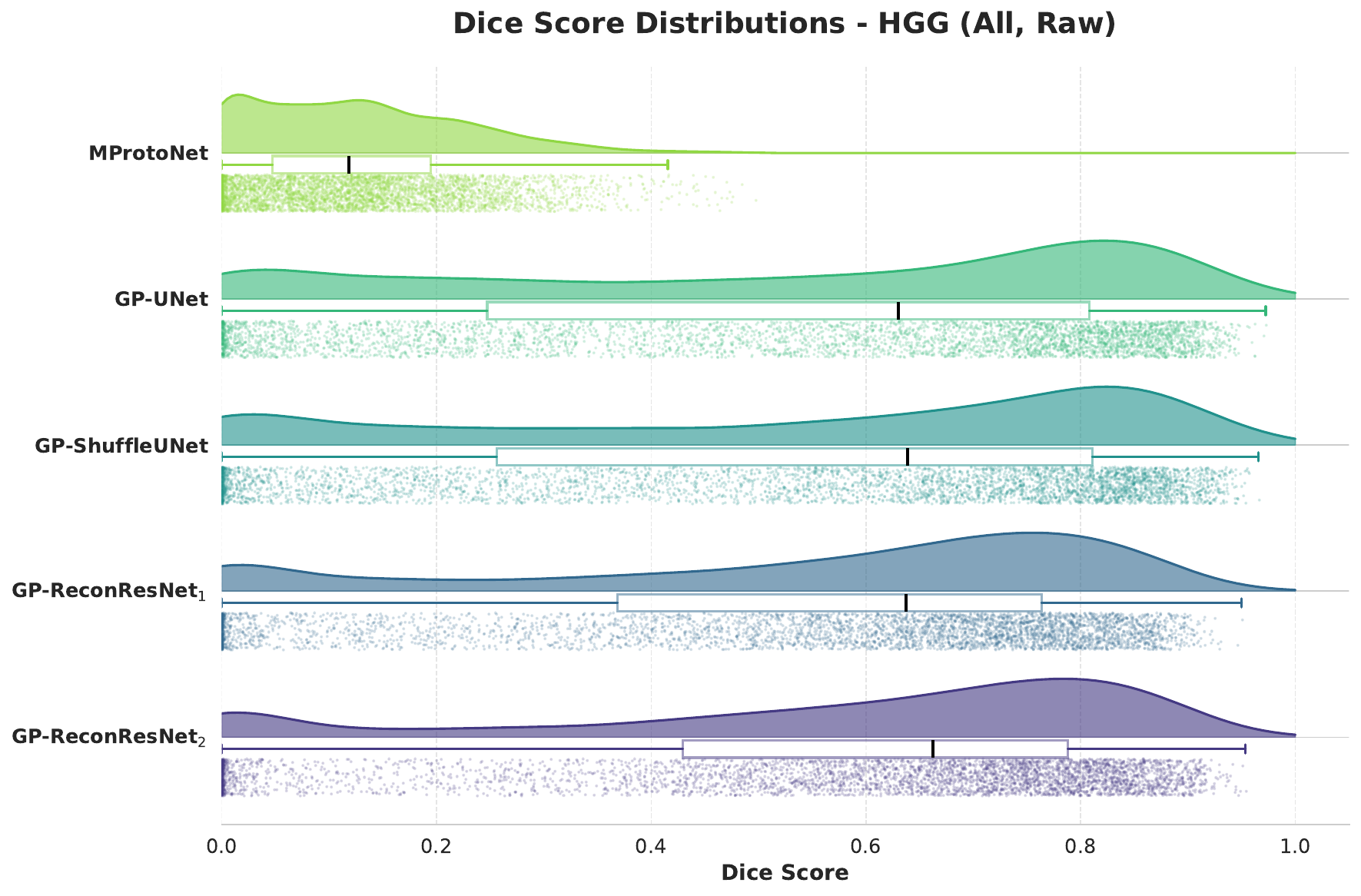}
    \caption{Stratified distribution of raw Dice scores for the HGG subtype across the full dataset. This final baseline confirms that the superior performance of the GP-ReconResNet models in this subtype is observable even in the raw, unfiltered model outputs.}
    \label{fig:app_hgg_all_raw}
\end{figure}

\FloatBarrier

\begin{figure}           %
\centering
\includegraphics[width=0.49\textwidth]{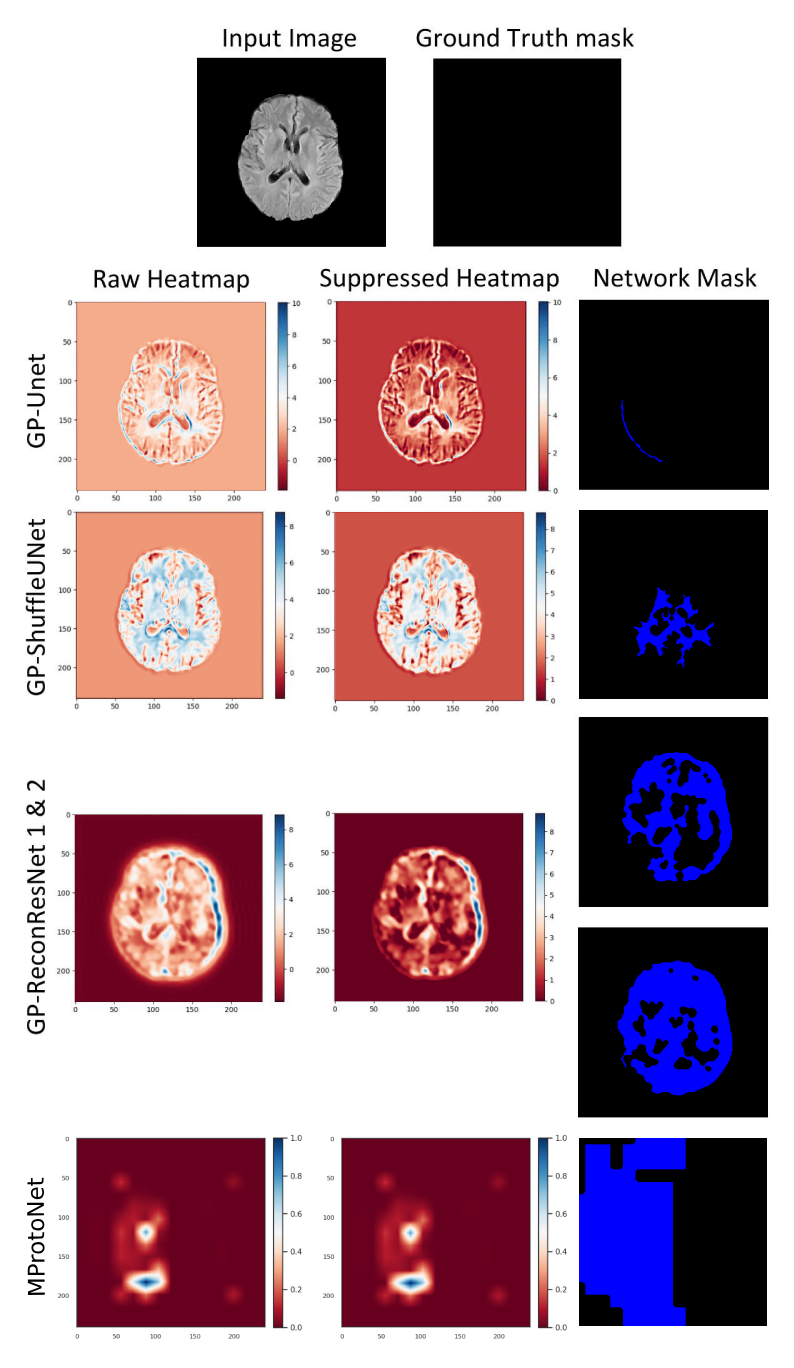}
\caption{Example results of the GP-models' correctly classified tumour-free slices (class 0) from the BraTS 2020 dataset. 1st row is the input slice, the ground truth mask is used for comparison, and the rest of the rows are the models' outputs. For the rest of the rows: 1st column contains the models' predictions known as the raw heatmaps, where the red areas influenced the classification outcome negatively, and the blue areas influenced the classification outcome favourably; 2nd column contains the suppressed heatmaps, where negative values are suppressed to obtain positive attributions only; 3rd column contains the networks' generated final masks = the suppressed heatmap + thresholding.}
\label{fig:MainResults1}
\end{figure}

\end{document}